\documentclass{aastex631}
\usepackage{chemfig}

\begin{document}

\title{Observations and chemical modeling of the isotopologues of formaldehyde and the cations of formyl and protonated formaldehyde in the hot molecular core G331.512-0.103}  

\correspondingauthor{Edgar Mendoza}
\email{edgar.mendoza@dci.uhu.es}

\author[0000-0001-9381-7826]{Edgar Mendoza}
\affiliation{Dept. Ciencias Integradas, Facultad de Ciencias Experimentales, Centro de Estudios Avanzados en F\'isica, Matem\'atica y Computaci\'on, Unidad Asociada GIFMAN, CSIC-UHU, Universidad de Huelva, Spain}

\author[0000-0001-8743-129X]{Miguel Carvajal}
\affiliation{Dept. Ciencias Integradas, Facultad de Ciencias Experimentales, Centro de Estudios Avanzados en F\'isica, Matem\'atica y Computaci\'on, Unidad Asociada GIFMAN, CSIC-UHU, Universidad de Huelva, Spain}
\affiliation{Instituto Universitario Carlos I de Física teórica y computacional, Universidad de Granada, Spain}

\author{Manuel Merello}
\affiliation{Departamento de Astronom{\'{\i}}a, Universidad de Chile, Casilla 36-D, Santiago de Chile, Chile}

\author{Leonardo Bronfman}
\affiliation{Departamento de Astronom{\'{\i}}a, Universidad de Chile, Casilla 36-D, Santiago de Chile, Chile}

\author{Heloisa M. Boechat-Roberty}
\affiliation{Observat\'orio do Valongo, Universidade Federal do Rio de Janeiro,  Ladeira do Pedro Ant\^onio, 43, Rio de Janeiro, RJ}



\begin{abstract}

In the interstellar cold gas, the chemistry of formaldehyde (H$_2$CO) can be essential to explain the formation of complex organic molecules. On this matter, the massive and energetic protostellar object G331 is still unexplored and, hence, we carried out a comprehensive study of the isotopologues of H$_2$CO and formyl cation (HCO$^+$), and of protonated formaldehyde (H$_2$COH$^+$) through the APEX observations in the spectral window $\sim$159--356~GHz.
We employed observational and theoretical methods to derive the physical properties of the molecular gas combining LTE and non-LTE analyses. Formaldehyde was characterized via 35 lines of H$_2$CO, H$_2^{13}$CO, HDCO and H$_2$C$^{18}$O. The formyl cation was detected via 8 lines of HCO$^+$, H$^{13}$CO$^+$, HC$^{18}$O$^+$ and HC$^{17}$O$^+$. Deuterium was clearly detected via HDCO, whereas DCO$^+$ remained undetected. H$_2$COH$^+$ was detected through 3 clean lines. 
According to the radiative analysis, formaldehyde appears to be embedded in a bulk gas with a wide range of temperatures ($T\sim$20--90 K), while HCO$^+$ and H$_2$COH$^+$ are primarily associated with a colder gas ($T\lesssim$ 30~K). The reaction H$_2$CO+HCO$^+ \rightarrow$ H$_2$COH$^+$ + CO is crucial for the balance of the three species. We used Nautilus gas-grain code to predict the evolution of their molecular abundances relative to H$_2$ which values at time scales $\sim$10$^3$~yr matched with the observations in G331: [H$_2$CO] = (0.2--2) $\times$10$^{-8}$, [HCO$^+$] = (0.5--4) $\times$10$^{-9}$ and [H$_2$COH$^+$] = (0.2--2) $\times$10$^{-10}$.
Based on the molecular evolution of H$_2$CO, HCO$^+$ and H$_2$COH$^+$, we hypothesized about the young lifetime of G331, which is consistent with the active gas-grain chemistry of massive protostellar objects.

\end{abstract}

\keywords{ISM: molecules --- astrochemistry ---  molecular data --- methods: data analysis --- line: identification}

\section{Introduction}\label{sec1}

Formaldehyde was one of the first organic molecules detected in the Interstellar Medium (ISM) \citep{Snyder1969,Gardner1974}. This compound has been observed both in the gas and solid phase of the ISM \citep{Meier1993,Schutte1996,Feraud2019}. Formaldehyde also plays a key role in the interstellar synthesis of pre-biotic and complex organic molecules  \citep{Ferus2019,Layssac2020,Paiva2023}.

In observational studies, \citet{Mangum1990} used maps and spectral lines of H$_2$CO to constrain the different components of the Orion Kleinmann–Low (KL) star forming region, which has been largely discussed in the context of cold, hot cores and molecular outflows \citep{Wootten1984,Sutton1995,Zapata2011}. \citet{Pegues2020} analyzed maps of H$_2$CO toward a sample of discs surveyed with ALMA. They highlighted the importance of H$_2$CO to understand the production mechanisms of O-bearing molecules in discs. In hot corinos and low-mass protostellar objects, H$_2$CO transitions have been useful to estimate its physical and chemical conditions \citep{Maret2004,Sahu2018,Martin2019}. In the Horsehead photodissociation region, \citet{Guzman2013} analyzed observations of H$_2$CO and CH$_3$OH to investigate their chemistry and dominant formation routes. In massive star forming regions, H$_2$CO transitions have been observed in hot molecular cores. In addition, an H$_2$CO maser at 6~cm has been proposed as an exclusive tracer of high-mass star formation \citep{Pratap1994,Araya2015}. In the IRAS 16562-3959 high-mass star-forming region, \citet{Taniguchi2020} observed H$_2$CO and investigate its formation pathways; they also analyzed the emission of (CH$_3$)$_2$CO and CH$_3$OCHO.  The identification of both simple and complex organic molecules contributes to our understanding of the chemical network of reactions that interplay these species under interstellar conditions (e.g. \citealt{Horn2004,Singh2022}). In evolved stellar objects, \citet{Ford2004} observed H$_2$CO toward the carbon star IRC+10216 and discussed the H$_2$CO formation in the context of solar system comet comae. In sources outside the Milky Way, \citet{Tang2021} observed H$_2$CO to construct a map of the kinetic temperature of two massive star-forming regions in the Large Magellanic Cloud (LMC).  \citet{Shimonishi2016} analyzed observations of  H$_2$CO, among other chemical species, to diagnose and claim the first detection of a hot molecular core in the LMC.

The physical conditions of the formaldehyde isotopologues are also investigated here. In a study about deuterated molecules in Orion KL, \citet{Neill2013} estimated D/H ratios from formaldehyde and other molecules using data from {\it Herschel}/HIFI. As a rare isotopic species, \citealt{Turner1990} detected D$_2$CO via four transitions toward Orion-KL. Nonetheless, \citet{Neill2013} did not confirm such detection. In a sample of Galactic molecular sources, \citet{Yan2019} performed a study about  the $^{12}$C/$^{13}$C ratio using transitions from the H$_2^{12}$CO and  H$_2^{13}$CO isotopologues. With respect to H$_2$C$^{18}$O, a few works have reported its detection. Using ALMA observations, \citet{Persson2018} not only detected H$_2$C$^{18}$O, but also H$_2$C$^{17}$O, D$_2^{13}$CO and HDC$^{18}$O, toward IRAS 16293-2422 B. 

\begin{figure}
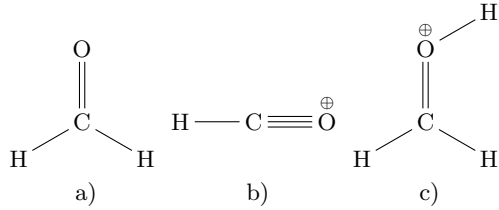

    \centering
     \chemname{\chemfig{C(-[:-150]H)(-[:-30]H)=[:90]O}}{a)}
     \chemname{\chemfig{H-C~\chemabove{O}{\scriptstyle \oplus}}}{b)}
     \chemname{\chemfig{C(-[:-150]H)(-[:-30]H)=[:90]\chemabove{O}{\scriptstyle\oplus}-[:30]H}}{c)}
\caption{Representation of the  neutral molecule a) formaldehyde, and the molecular ions b) formyl cation and c) protonated formaldehyde.}
\label{fig:fig1}
\end{figure}

In chemical association with H$_2$CO, we also present results about the molecular ions HCO$^+$ and H$_2$COH$^+$  (Fig.~\ref{fig:fig1}). HCO$^+$ and its isotopologues have been observed in hot molecular cores and outflows \citep{Sanchez2013,Shimonishi2016}. In previous works, \citet{mer2013b} and \citet{hervias2019} reported preliminary results about H$^{13}$CO$^+$ in G331.512-0.103, from now on G331.
H$_2$COH$^+$ is a rare molecular ion whose detection has been scarcely reported in the literature. It has been observed toward Sgr B2, Orion KL, W51 and the prestellar core L1689B \citep{ohi1996,bac2016}.

H$_2$CO, HCO$^+$ and H$_2$COH$^+$ are chemical species that participate and compete in various chemical reactions, for instance:  

\begin{equation} \label{Eq:1}
\text{H$_2$CO + HCO$^+ \longrightarrow$ H$_2$COH$^+$ + CO ~~~.} 
\end{equation}

\noindent The kinetic constants were experimentally studied in an early work by \citet{Tanner1979}. \citet{bac2016} observed H$_2$COH$^+$ in the prestellar source L1689B and discussed the reaction presented in Eq.~\ref{Eq:1} in the context of cold prestellar cores. The three species studied here, H$_2$CO, HCO$^+$ and H$_2$COH$^+$, also play a key role in the gas-grain chemistry to form complex organic molecules, such as glycolaldehyde and sugar related molecules \citep{hal2006,woo2012,eck2018,Layssac2020}.  Additionally, it is worth mentioning that H$_2$CO is closely associated with methanol (CH$_3$OH), a pivotal molecule for interstellar chemical complexity. Studies have demonstrated that successive hydrogenation reactions of carbon monoxide (CO) can lead to the formation of both H$_2$CO and CH$_3$OH \citep{Watanabe2003,Tsuge2020}.

The present work is based on observations of G331, a massive protostellar object embedded in the G331.5-0.1 giant molecular cloud.  The central object drives a powerful outflow with a flow mass and momentum of $\sim 55 ~M_\odot$ and $\sim 2.4 \times 10^3 ~M_\odot$ km s$^{-1}$, respectively \citep{Bronfman2008}. The source is located in the tangent region of the Norma spiral arm, at an heliocentric distance of $\sim$~7.5~kpc. Regarding the ambient core of G331, ALMA observations evidence a lukewarm gas with $n_{\text{H}_2} \sim 5 \times 10^6$~cm$^{-3}$ and $T \sim$ 70 K \citep{hervias2019}. In association with those gas conditions, \citet{Canelo2021} and \citet{Santos2022} reported recent results about the physical and chemical conditions of the isocyanic acid (HNCO) and methyl acetylene (CH$_3$CCH). Here, we then contribute with new results about the lukewarm and cold gas conditions of G331, which are based on a broad spectral survey collected with the Atacama Pathfinder Experiment (APEX). 

In massive protostellar objects, the chemical scenario about the formation of hydrocarbons, carbon chain and organic compounds is not well understood yet \citep{Taniguchi2018,Kalvans2021}. In view of the importance to understand how gas-grain processes occur in massive protostellar objects, this work also presents a chemical model to understand the cold chemistry that connects two abundant species, as H$_2$CO and HCO$^+$, with a low abundant ion as H$_2$COH$^+$. In addition, this study allowed us to venture the age of the protostellar object G331.
This article is organized as follows: Section~\ref{sec2} describes the observations carried out by APEX of the source G331 and the methodology used for this study. Section~\ref{sec3} presents the results of the spectral analysis and the estimated physical conditions of G331. In Section~\ref{sec4}, the astrophysical and astrochemical implications of the present results are discussed. Final remarks and conclusions are summarized in Section~\ref{sec5}. Additionally, an appendix section (\ref{appendix}) has been included to provide supplementary information and data.

\section{Methodology} \label{sec2}

The APEX telescope \citep{gu06} was used to perform the observations, adopting the single point mode, toward the source coordinates RA:DEC = 16$^h$12$^m$10.1$^s$, $-$51$^{\circ}$28$^{\prime}$38.1$^{\prime\prime}$. The APEX-1 and APEX-2 receivers, of the Swedish Heterodyne Facility Instrument (SHeFI)  \citep{vas08}, were used to collect spectral setups in the frequency ranges 213--275~GHz and 267--378~GHz, respectively. The spectral resolution of the dataset was adjusted to be between $\sim$0.15 and 0.25 km s$^{-1}$  for a noise level of about 30~mK  across the bands. 

As part of the observational runs, the SEPIA B5 instrument \citep{belitsky2018} was used to collect setups between 159--211~GHz.  SEPIA B5 is a dual-polarisation sideband-separated receiver. The  lower  and  upper  sideband (LSB  and  USB)  are  separated  by  12  GHz. Each  sideband is recorded by two XFFTS units of 2.5 GHz with a 1 GHz overlap. The Half Power Beam Width (HPBW) values, covered by the APEX receivers, are ranged from $\sim$~17 to 39\arcsec, were estimated for each transition using 7.8\arcsec $\times$ 800/$\nu$[GHz] \citep{gu06,Quenard2017}\footnote{\url{https://www.apex-telescope.org/ns/instruments/}}, where $\nu$ is the rest frequency of the spectra. With respect to the calibration uncertainty, \citet{dum2010} discussed that lines of C$^{18}$O (3–2) and $^{13}$CO (3–2) might reach uncertainties of $\sim$ 13~\%, and lines of H$_2$CO and CH$_3$OH of up to $\sim$~33 \%. Considering that we observed different species through a multi-wavelength analysis, it was adopted an overall calibration uncertainty of  30~\%.

The data reduction was carried out with the CLASS package of GILDAS.\footnote{\url{https://www.iram.fr/IRAMFR/GILDAS/}} The spectra obtained from the APEX real-time calibration tool \citep{Muders2006}, in the CLASS data architecture, are in the corrected antenna temperature scale.\footnote{\url{https://www.apex-telescope.org/telescope/efficiency/}} For further clarity, we exhibit spectra in units of antenna temperature (K) with respect to the systemic velocity of G331 ($V_{\rm lsr}$ = -90 km s$^{-1}$). Additionally, all the spectra were smoothed to exhibit a common channel of~1~km~s$^{-1}$.
The Weeds\footnote{\url{https://www.iram.fr/IRAMFR/GILDAS/doc/html/weeds-html/weeds.html}} extension of CLASS \citep{Maret2011}, along with the spectroscopic databases CDMS\footnote{\url{https://cdms.astro.uni-koeln.de/}} \citep{end2016} and 
JPL\footnote{\url{https://spec.jpl.nasa.gov/}} \citep{pic1998}, were also used here.
The CASSIS\footnote{\url{http://cassis.irap.omp.eu/}} software was used to estimate the physical conditions from the spectral lines, which were utilized in the main beam temperature ($T_{mb}$) scale adopting the main-beam efficiencies $\eta_{mb}$~= 0.80, 0.75 and 0.73 for SEPIA180, APEX-1 and APEX-2, respectively. In order to evaluate beam dilution effects, a source size of 5\arcsec \ was adopted for the G331 core, and of 15\arcsec \ for an expanded region. Using the CASSIS tools, Local Thermodynamic Equilibrium analyses were performed using calculations and the population diagram method \citep{Goldsmith1999,Mangum2015,Roueff2021}. Non-LTE calculations were carried out with the RADEX code \citep{van2007} using rate coefficients from the LAMDA database.\footnote{\url{http://home.strw.leidenuniv.nl/~moldata/}} 

\subsection{Chemical models}\label{sec2.1}
The gas grain code \textsc{Nautilus} was used to compute abundances as a function of time from a network of gas and grain chemical reactions  \citep{sem10,reb14,rua15,nautilus}. We focused on computing the abundances of H$_2$CO, HCO$^+$ and H$_2$COH$^+$ considering similar conditions to those obtained from the observations. \textsc{Nautilus} uses the KIDA\footnote{\url{https://kida.astrochem-tools.org/}} database \citep{wak15}, which includes rate coefficients for a total of about 7509 reactions and 489 chemical species. For the solid state chemistry, it considers mantle and surface mechanisms as those investigated by \citet{has93} and \citet{ghe15}. For the neutral and molecular ions analyzed in this work, there is a network of no more than 50 chemical reactions connecting them. The chemistry of HCO$^+$ and H$_2$CO is significant and the detection of various of their isotopologues suggests gas-grain processes in cold environments. Previous studies unveiled a rich chemistry in  lukewarm regions of G331 ($T_g \thickapprox$70~K) evidenced from various spectral lines of HNCO and CH$_3$CCH \citep{Canelo2021,Santos2022}. We expanded here the chemical models of G331 adopting a cold gas condition ($T_g \thickapprox$50~K), which might not only explain the abundances of H$_2$CO, HCO$^+$ and H$_2$COH$^+$ but also the eventual formation  of more complex molecules.

\section{Results} \label{sec3}

\subsection{Line analysis} \label{sec3.1}

\subsubsection{Formaldehyde and its isotopologues} \label{sec3.1.1}

\begin{figure}
\centering
\includegraphics[width=14cm,keepaspectratio]{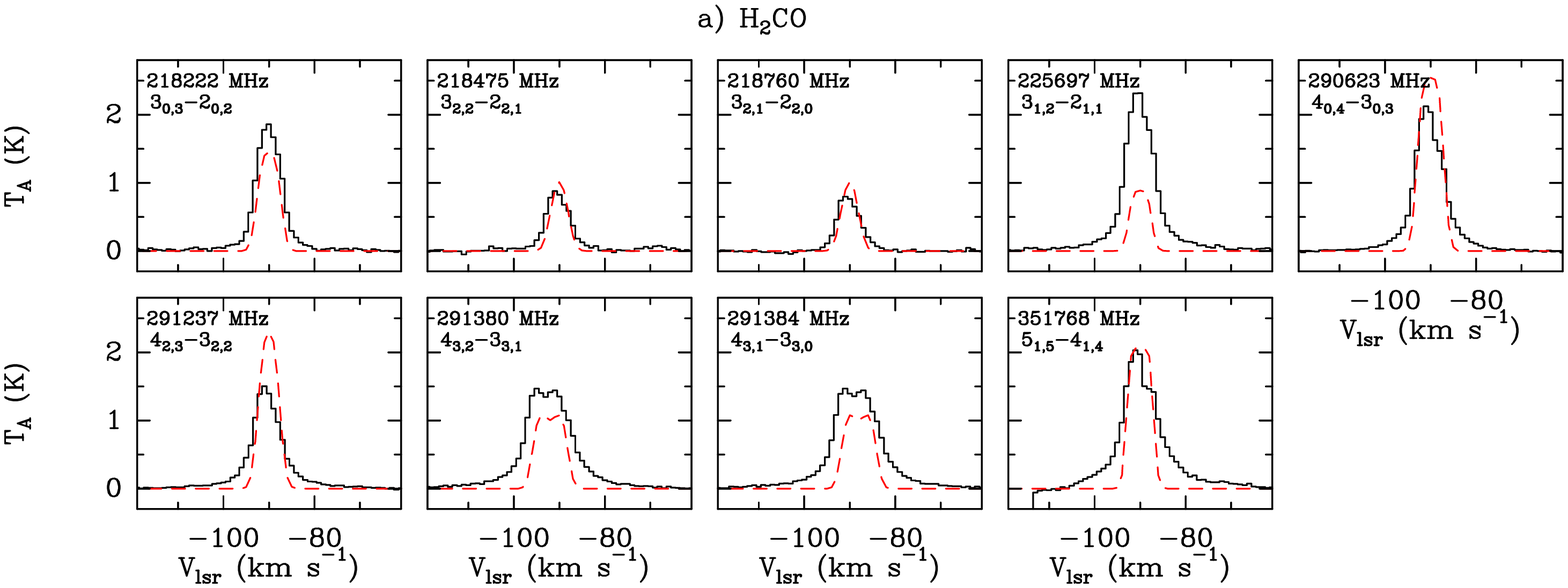}\\
\includegraphics[width=14cm,keepaspectratio]{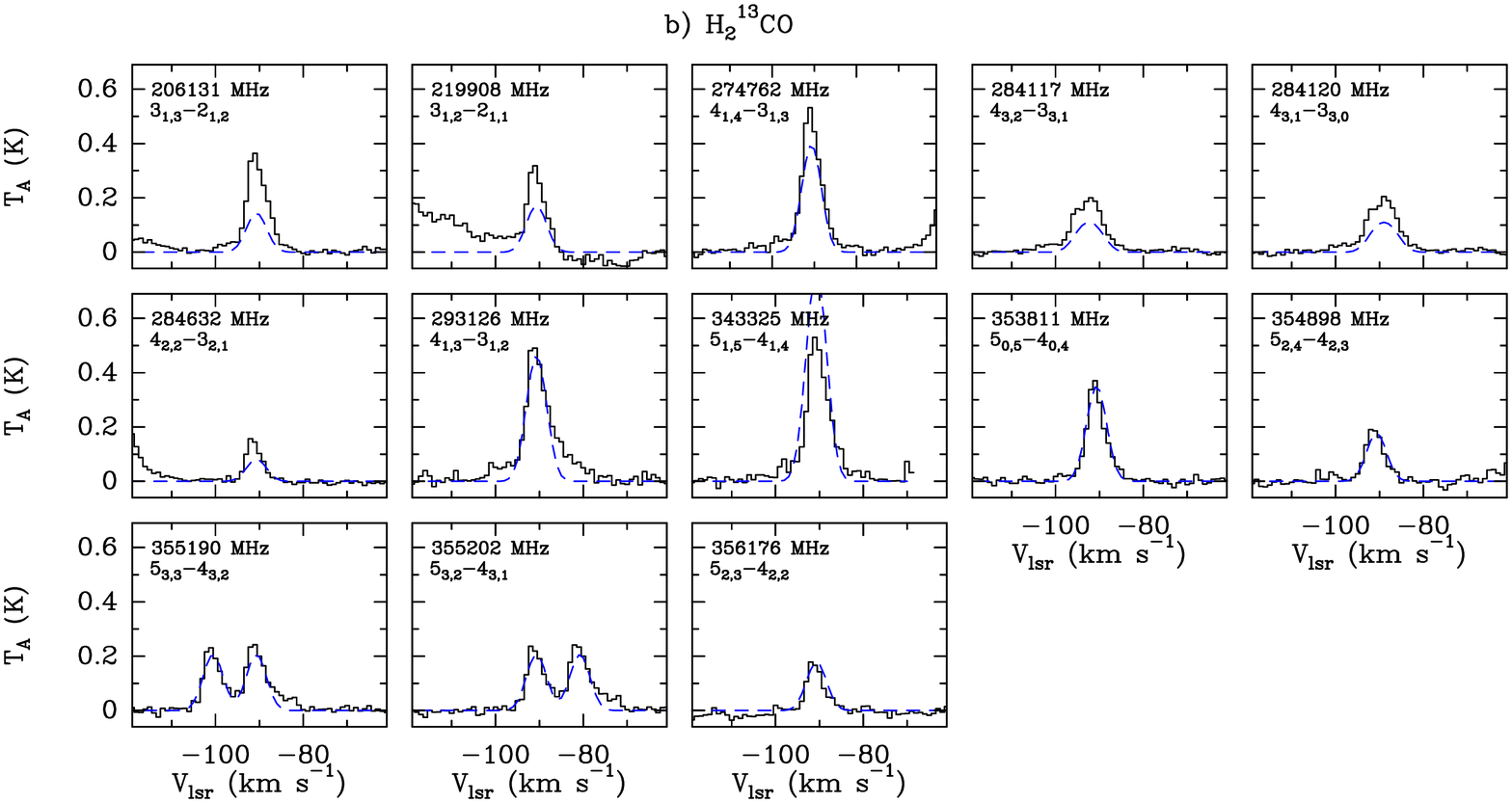}\\
\includegraphics[width=14cm,keepaspectratio]{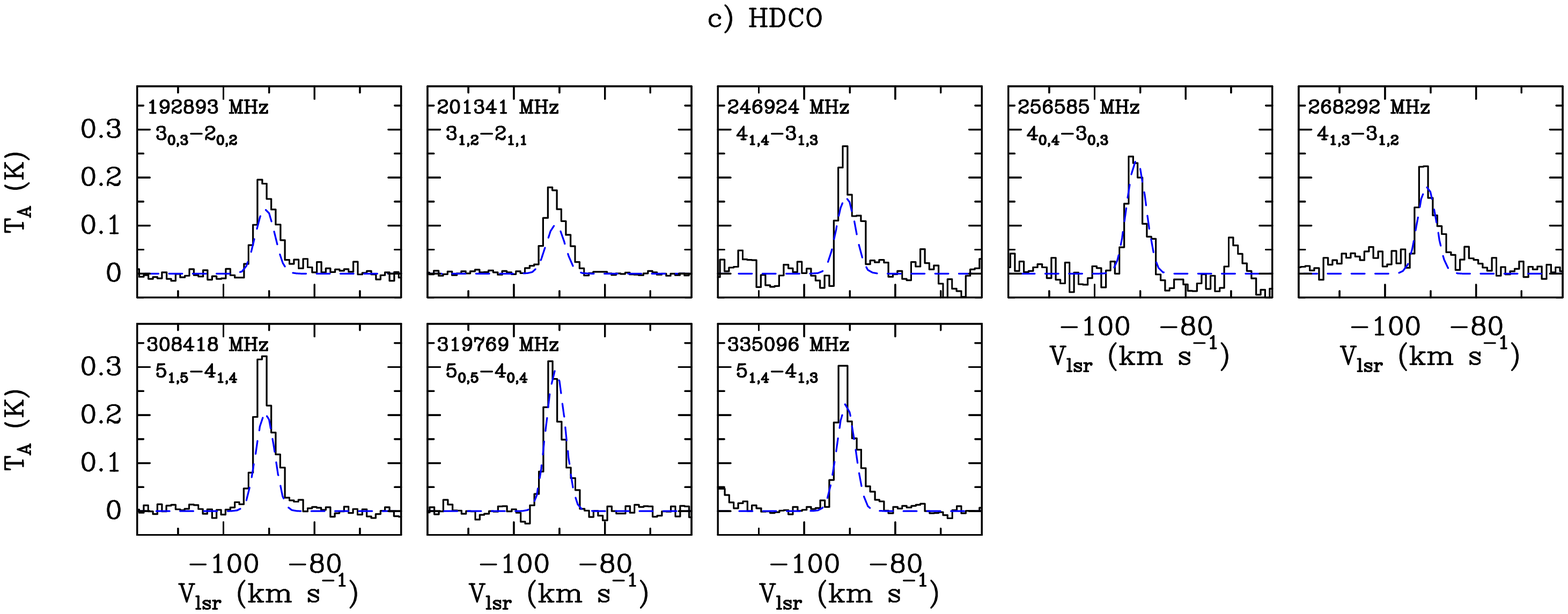}\\
\includegraphics[width=14cm,keepaspectratio]{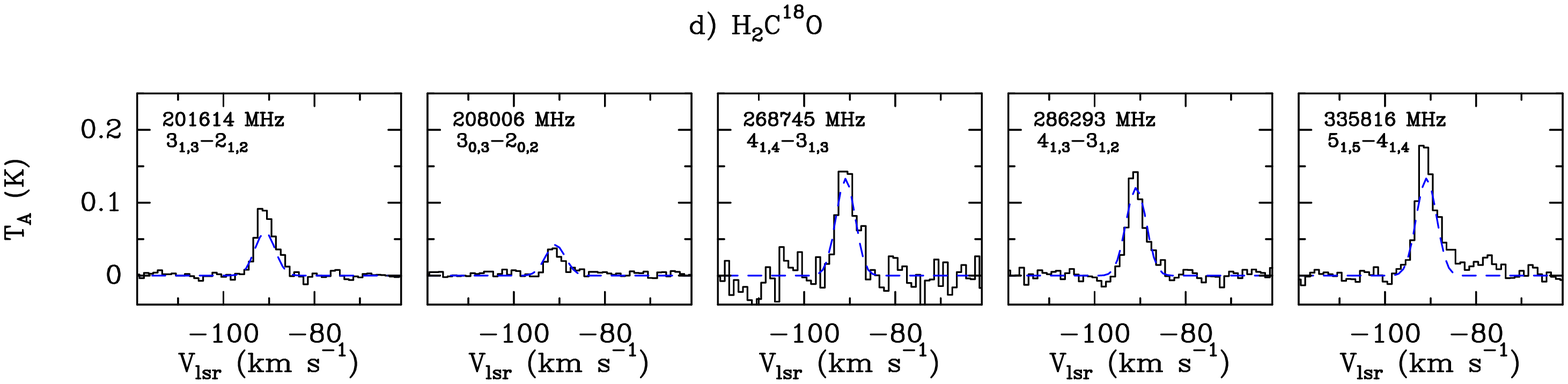}
\caption{Observed (solid) and modelled (dashed) spectra, from the LTE  (in blue) and non-LTE  (in red) simulations (\S~\ref{sec3.2}), of the formaldehyde isotopologues in G331: a) H$_2$CO, b) H$_2^{13}$CO, c) HDCO and d) H$_2$C$^{18}$O. For each spectrum,  the $y$-axis represents the intensity in a scale of antenna temperature (K) while the $x$-axis represents the velocity relative to the local standard of rest (Systemic $V_{\rm lsr}=$ -90~km~s$^{-1}$).}
\label{fig:fig2} 
\end{figure}    

\begin{table}
\centering
\caption{Spectral line analysis of H$_2^{12}$CO, H$_2^{13}$CO, HDCO and H$_2$C$^{18}$O.}
\label{tab:tab1}
\begin{tabular}{lllccccccc}
\hline\hline 
Species and transitions & Frequency\tablenotemark{a} & Beamwidth & $E_u$ & $g_{u}$ & $A_{ul}$ & Line area  &  $V_{\rm lsr}$ & Linewidth & rms \\
             & [MHz]     & [\arcsec] & [K] & & [10$^{-5}$s$^{-1}$] & [K km s$^{-1}$] &  [km s$^{-1}$] & [km s$^{-1}$]   & [mK] \\
\hline
p-H$_2$CO	(3	0	3	--	2	0	2)	&	218222.192	&	28.56	&	20.96	&	7	&	28.2	&	12.99	$\pm$	0.07	&	-90.07	$\pm$	0.02	&	6.4	$\pm$	0.04	&	47	\\
p-H$_2$CO	(3	2	2	--	2	2	1)	&	218475.632	&	28.56	&	68.09	&	7	&	15.7	&	5.5	$\pm$	0.3	&	-90.4	$\pm$	0.1	&	5.9	$\pm$	0.3	&	35	\\
p-H$_2$CO	(3	2	1	--	2	2	0)	&	218760.066	&	28.52	&	68.11	&	7	&	15.8	&	5.04	$\pm$	0.04	&	-90.33	$\pm$	0.02	&	5.85	$\pm$	0.05	&	36	\\
o-H$_2$CO	(3	1	2	--	2	1	1)	&	225697.775	&	27.65	&	33.45	&	21	&	27.7	&	17.88	$\pm$	0.06	&	-90.2	$\pm$	0.01	&	7.25	$\pm$	0.03	&	38	\\
p-H$_2$CO	(4	0	4	--	3	0	3)	&	290623.405	&	21.47	&	34.9	&	9	&	69	&	16.1	$\pm$	0.1	&	-90.39	$\pm$	0.03	&	7.29	$\pm$	0.08	&	42	\\
p-H$_2$CO	(4	2	3	--	3	2	2)	&	291237.78	&	21.43	&	82.07	&	9	&	52.1	&	10.92	$\pm$	0.03	&	-90.578	$\pm$	0.009	&	7.13	$\pm$	0.02	&	33	\\
o-H$_2$CO	(4	3	2	--	3	3	1)	&	291380.488	&	21.41	&	140.94	&	27	&	30.4	&	9.653	$\pm$	1E-3	&	-90.36	$\pm$	0.02	&	6.73	$\pm$	0.04	&	28	\\
o-H$_2$CO	(4	3	1	--	3	3	0)	&	291384.264	&	21.41	&	140.94	&	27	&	30.4	&	12	$\pm$	0.1	&	-90.38	$\pm$	0.04	&	7.75	$\pm$	0.09	&	29	\\
o-H$_2$CO	(5	1	5	--	4	1	4)	&	351768.645	&	17.74	&	62.45	&	33	&	120	&	14.7	$\pm$	0.2	&	-90.88	$\pm$	0.04	&	7.1	$\pm$	0.1	&	46\\
\hline
o-H$_2^{13}$CO	(3	1	3	--	2	1	2)	&	206131.626	&	30.27	&	31.62	&	21	&	21.1	&	1.68	$\pm$	0.02	&	-90.91	$\pm$	0.03	&	4.37	$\pm$	0.08	&	18	\\
o-H$_2^{13}$CO	(3	1	2	--	2	1	1)	&	219908.525	&	28.37	&	32.94	&	21	&	25.6	&	1.14	$\pm$	0.04	&	-91.01	$\pm$	0.06	&	3.6	$\pm$	0.2	&	34	\\
o-H$_2^{13}$CO	(4	1	4	--	3	1	3)	&	274762.112	&	22.71	&	44.8	&	27	&	54.7	&	2.49	$\pm$	0.04	&	-90.94	$\pm$	0.03	&	4.47	$\pm$	0.08	&	32	\\
o-H$_2^{13}$CO	(4	3	2	--	3	3	1)	&	284117.45	&	21.96	&	140.47	&	27	&	28.2	&	1.55	$\pm$	0.05	&	-92.1	$\pm$	0.1	&	7.3	$\pm$	0.3	&	21	\\
o-H$_2^{13}$CO	(4	3	1	--	3	3	0)	&	284120.62	&	21.96	&	140.47	&	27	&	28.2	&	1.1	$\pm$	0.1	&	-90.5	$\pm$	0.4	&	6.3	$\pm$	0.7	&	 18	\\
p-H$_2^{13}$CO	(4	2	2	--	3	2	1)	&	284632.42	&	21.92	&	81.43	&	9	&	48.6	&	0.59	$\pm$	0.02	&	-91.54	$\pm$	0.05	&	3.4	$\pm$	0.1	&	17	\\
o-H$_2^{13}$CO	(4	1	3	--	3	1	2)	&	293126.515	&	21.29	&	47.01	&	27	&	66.4	&	2.25	$\pm$	0.08	&	-91.2	$\pm$	0.07	&	4.2	$\pm$	0.2	&	42	\\
o-H$_2^{13}$CO	(5	1	5	--	4	1	4)	&	343325.713	&	18.17	&	61.28	&	33	&	112	&	2.45	$\pm$	0.09	&	-90.93	$\pm$	0.07	&	4.3	$\pm$	0.2	&	30	\\
p-H$_2^{13}$CO	(5	0	5	--	4	0	4)	&	353811.872	&	17.64	&	51.02	&	11	&	127	&	1.54	$\pm$	0.04	&	-91.22	$\pm$	0.05	&	3.8	$\pm$	0.1	&	33	\\
p-H$_2^{13}$CO	(5	2	4	--	4	2	3)	&	354898.595	&	17.58	&	98.41	&	11	&	108	&	0.83	$\pm$	0.05	&	-91.3	$\pm$	0.1	&	4.0	$\pm$	0.3	&	34	\\
o-H$_2^{13}$CO	(5	3	3	--	4	3	2)	&	355190.9	&	17.57	&	157.52	&	33	&	82.5	&	1.39	$\pm$	0.04	&	-90.81	$\pm$	0.08	&	5.9	$\pm$	0.2	&	29	\\
o-H$_2^{13}$CO	(5	3	2	--	4	3	1)	&	355202.601	&	17.57	&	157.52	&	33	&	82.5	&	1.21	$\pm$	0.03	&	-90.79	$\pm$	0.07	&	5.1	$\pm$	0.2	&	32	\\
p-H$_2^{13}$CO	(5	2	3	--	4	2	2)	&	356176.243	&	17.52	&	98.52	&	11	&	109	&	1.02	$\pm$	0.03	&	-91.04	$\pm$	0.08	&	5.2	$\pm$	0.2	&	32	\\
\hline
HDCO	(3	0	3	--	2	0	2)	&	192893.27	&	32.35	&	18.53	&	7	&	19.4	&	1.03	$\pm$	0.02	&	-90.73	$\pm$	0.06	&	5.2	$\pm$	0.1	&	21	\\
HDCO	(3	1	2	--	2	1	1)	&	201341.35	&	30.99	&	27.29	&	7	&	19.6	&	0.956	$\pm$	0.008	&	-91.02	$\pm$	0.02	&	5.25	$\pm$	0.05	&	18	\\
HDCO	(4	1	4	--	3	1	3)	&	246924.6	&	25.27	&	37.6	&	9	&	39.6	&	1.06	$\pm$	0.04	&	-90.73	$\pm$	0.09	&	4.3	$\pm$	0.2	&	47	\\
HDCO	(4	0	4	--	3	0	3)	&	256585.43	&	24.32	&	30.85	&	9	&	47.4	&	1.25	$\pm$	0.05	&	-91.11	$\pm$	0.08	&	4.6	$\pm$	0.2	&	50	\\
HDCO	(4	1	3	--	3	1	2)	&	268292.02	&	23.26	&	40.17	&	9	&	50.8	&	0.85	$\pm$	0.03	&	-91.06	$\pm$	0.09	&	4.1	$\pm$	0.2	&	37	\\
HDCO	(5	1	5	--	4	1	4)	&	308418.2	&	20.23	&	52.4	&	11	&	80.8	&	1.54	$\pm$	0.02	&	-91.14	$\pm$	0.02	&	4.72	$\pm$	0.07	&	20	\\
HDCO	(5	0	5	--	4	0	4)	&	319769.68	&	19.51	&	46.19	&	11	&	93.7	&	1.41	$\pm$	0.03	&	-91.31	$\pm$	0.05	&	4.5	$\pm$	0.1	&	37	\\
HDCO	(5	1	4	--	4	1	3)	&	335096.739	&	18.62	&	56.25	&	11	&	104	&	1.43	$\pm$	0.02	&	-91.04	$\pm$	0.04	&	4.76	$\pm$	0.09	&	28	\\
\hline
o-H$_2$C$^{18}$O	(3	1	3	--	2	1	2)	&	201614.256	&	30.95	&	31.22	&	21	&	19.7	&	0.473	$\pm$	0.009	&	-90.83	$\pm$	0.05	&	4.8	$\pm$	0.1	&	18	\\
p-H$_2$C$^{18}$O	(3	0	3	--	2	0	2)	&	208006.441	&	29.99	&	19.97	&	7	&	24.4	&	0.134	$\pm$	0.007	&	-91.17	$\pm$	0.09	&	3.4	$\pm$	0.2	&	17	\\
o-H$_2$C$^{18}$O	(4	1	4	--	3	1	3)	&	268745.789	&	23.22	&	44.12	&	27	&	51.1	&	0.76	$\pm$	0.04	&	-90.7	$\pm$	0.1	&	4.9	$\pm$	0.3	&	48	\\
o-H$_2$C$^{18}$O	(4	1	3	--	3	1	2)	&	286293.96	&	21.79	&	46.23	&	27	&	61.8	&	0.64	$\pm$	0.02	&	-91.01	$\pm$	0.06	&	4.3	$\pm$	0.2	&	23	\\
o-H$_2$C$^{18}$O	(5	1	5	--	4	1	4)	&	335816.025	&	18.58	&	60.24	&	33	&	105	&	0.85	$\pm$	0.02	&	-90.99	$\pm$	0.06	&	4.8	$\pm$	0.1	&	24	\\
\hline
\end{tabular}
\tablenotetext{a}{Rest frequency values, obtained from databases such as CDMS \citep{end2016} and JPL \citep{pic1998}, are accessed using the CASSIS software \citep{Vastel2015}(see \S~\ref{sec2} for more details)}.
\end{table}

The isotopologues H$_2^{12}$CO, H$_2^{13}$CO, HDCO and H$_2$C$^{18}$O were identified in the frequency interval  $\sim$~190--357~GHz through 35 transition lines whose assignments, specifying the ortho or para character of the states for the symmetric isotopologues (e.g. \citealt{Clouthier1983,CHAPOVSKY2001}), spectroscopic data and Gaussian fit parameters are given in Table~\ref{tab:tab1}. The spectral analysis was primarily performed on lines with significantly stronger signals than the limit of detection (3$\sigma$ level), which are exhibited in the different panels of Fig.~\ref{fig:fig2}. In agreement with the expected isotopic abundances, the most and least abundant isotopologues, H$_2^{12}$CO and H$_2$C$^{18}$O, respectively, exhibited the strongest and weakest line intensities, respectively. Despite H$_2$C$^{18}$O is a rare isotopologue (e.g., \citealt{Muller2017}), we could detect it via 5 spectral lines. 

The main isotopologue (o, p)-H$_2$CO was detected via 9 lines although the lines at $\sim$~291380.48 and 291384.26~MHz, corresponding to the transitions 4$_{3,2}$-3$_{3,1}$ and 4$_{3,1}$-3$_{3,0}$, respectively, can be considered partially resolved (see Fig~\ref{fig:fig2}(a)). Various of the H$_2$CO transitions detected in this work were also reported in sources as OMC-1, Orion-KL and in high-mass star-forming regions   (e.g. \citealt{Loren1984,Wootten1984,Mangum1990,Taniguchi2020}). 

13 spectral lines of (o,~p)-H$_2^{13}$CO were identified although two of them at the rest frequencies $\sim$~284117.45 and 284120.62~MHz were deemed as partially resolved (see Fig~\ref{fig:fig2}(b)).  As it is expected, the intensities of the H$_2^{13}$CO lines are lower than those of H$_2$CO. By comparing the integrated areas of neighbor lines, e.g., o-H$_2^{13}$CO  at $\sim$~219908.52~MHz and o-H$_2$CO at $\sim$~225697.77~MHz, the H$_2$CO/H$_2^{13}$CO ratio is about 16. In previous works, \citet{Jewell1989} and \citet{Helmich1997} observed various of the H$_2^{13}$CO transitions in G331. 

The HDCO isotopologue was detected via 8 spectral  lines (see Fig.~\ref{fig:fig2}(c)).
It can be noted that the HDCO line profiles exhibit a partial asymmetry with a blue-shifted emission wing. In an investigation on HNCO in G331, \citet{Canelo2021} not only observed similar spectral asymmetries but also found them to be more pronounced in some specific $K$-ladder transitions of HNCO. These findings were discussed in the context of molecular outflows  (e.g. \citealt{Canelo2021} and references therein). In this work, we expect that the HDCO emission may be linked to an expanded gas region influenced by the molecular outflow. However, to better understand the emission of spectral tracers potentially associated with the molecular outflow, it is crucial to conduct further investigations, including the development of models that consider the core and outflow of G331.

In addition, 5 spectral lines of (o, p)-H$_2$C$^{18}$O isotopologue were clearly identified and shown in Fig.~\ref{fig:fig2}(d). From this study toward G331, we
confirm the observation of the transitions of the rare isotopologue H$_2$C$^{18}$O previously reported toward other sources (e.g., \citealt{Mangum1990,Sutton1995}). 

\begin{figure}
\centering
\includegraphics[width=8.5cm,keepaspectratio]{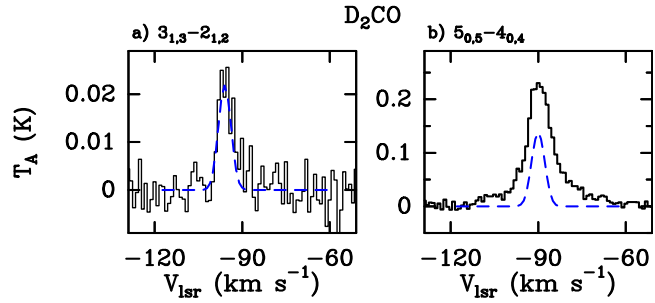}
\caption{Tentative identification of D$_2$CO: a) p-D$_2$CO 3$_{1,3}$--2$_{1,2}$ at the frequency $\sim$~166102.74~MHz and appearing at $V_{\rm lsr} \thickapprox -96$ km~s$^{-1}$; b) o-D$_2$CO 5$_{0,5}$--4$_{0,4}$ at the frequency $\sim$~287485.65~MHz but likely dominated by emission of SO$_2$ at 287485.44~MHz. Under the upper limit condition $N$(D$_2$CO)$<N$(HDCO), and FWHM=5~km~s$^{-1}$, the dashed lines indicate hypothetical LTE models of D$_2$CO.}
\label{fig:fig3}
\end{figure}

The successful detection of HDCO prompted further investigation into the presence of the doubly deuterated formaldehyde (D$_2$CO), which has been scarcely observed in objects of the ISM (e.g., \citealt{Turner1990,Ceccarelli1998}). As a result, we report the tentative detection of p-D$_2$CO 3$_{1,3}$--2$_{1,2}$ at the rest frequency $\sim$~166102.74~MHz previously detected  in prestellar cores \citep{Bacmann2003}. Fig.~\ref{fig:fig3}(a) shows this spectral line whose emission is above the limit of detection although shifted from the source systemic velocity ($V_{\rm lsr} = -90$~km~s$^{-1}$).  This makes us think that a possible candidate for this line might also be OC$^{34}$S (14--13) at 166105.75~MHz. A second tentative identification is displayed in  Fig.~\ref{fig:fig3}b,  the o-D$_2$CO 5$_{0,5}$--4$_{0,4}$ transition was also tentatively identified but a dominant line, likely blended with SO$_2$ v=0 at $\sim$~287485.44~MHz, avoided a clear identification.

\subsubsection{The formyl cation and its isotopologues}  \label{sec3.1.2}
    
\begin{figure}
\centering
\includegraphics[width=8.5cm,keepaspectratio]{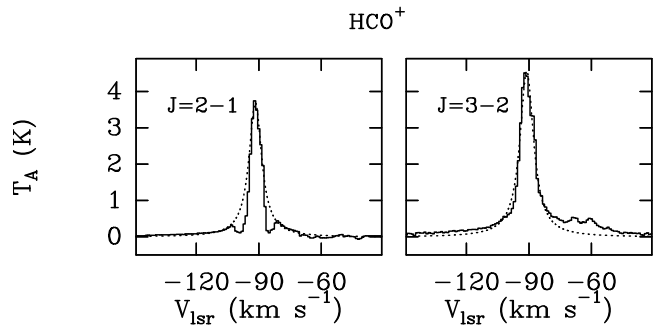}
\caption{Spectral lines (solid lines) and Lorentzian fits (dotted lines) of the transitions 2--1 and 3--2 of HCO$^+$ identified at the rest frequencies~$\sim$~178375.05~MHz and 267557.62~MHz, respectively.}
\label{fig:fig4}
\end{figure}

The isotopologues of formyl cation (HCO$^+$, H$^{13}$CO$^+$, DCO$^+$, HC$^{18}$O$^+$ and HC$^{17}$O$^+$) were also sought in G331. Eight lines from these isotopologues were identified, except the  deuterated one DCO$^+$, of which their spectroscopic and fitted parameters are given in Table~\ref{tab:tab2}.
The two rotational lines of the main isotopologue HCO$^+$ and their fits are shown in  Fig.~\ref{fig:fig4}. The HCO$^+$ spectra exhibited broad spectral wings, from $-$150~km~s$^{-1}$ to $-$30~km~s$^{-1}$, and they were better described by Lorentzian than Gaussian functions. The Lorentzian profile suggests that the HCO$^+$ emission is likely affected by the molecular outflow \citep{hervias2019}. 

\begin{figure}
\centering
\includegraphics[width=8.5cm,keepaspectratio]{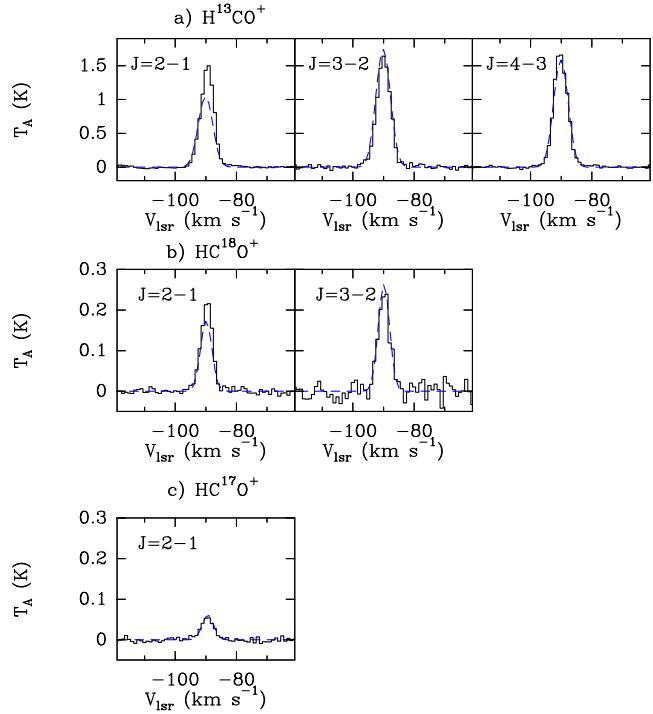}
\caption{Legend similar to that of Fig.~\ref{fig:fig2} but for the isotopologues of the formyl cation, with their LTE models: a) H$^{13}$CO$^+$, b) HC$^{18}$O$^+$ and c)  HC$^{17}$O$^+$.}
\label{fig:fig5}
\end{figure}    

The spectral lines of H$^{13}$CO$^+$, HC$^{18}$O$^+$ and HC$^{17}$O$^+$ did not exhibit Lorentzian profiles with broad wings. Thus, Gaussian functions, instead of Lorentzian ones, were used to fit the spectra. The spectra of these isotopologues are exhibited in the different panels of Fig.~\ref{fig:fig5}. For all the identified HCO$^+$ isotopologues, the $J$=2--1 transition was observed. Considering the velocity integrated temperatures for the line profiles of this transition,  the ratios  HCO$^+$:H$^{13}$CO$^+$:HC$^{18}$O$^+$:HC$^{17}$O$^+ \thickapprox$ 64:21:3:1 were obtained. Nevertheless, the results based on LTE and non-LTE methods will be given in the next section.

It is worth highlighting that the deuterated formyl cation DCO$^+$ was not detected in spite of the isotopologue HDCO was identified through several lines. In contrast, the $^{17}$O isotopologue of formyl cation was detected but not the isotopologue of formaldehyde H$_2$C$^{17}$O. Those aspects demand follow-up studies due to their implications on the understanding of the isotopic fraction and the evolution of protostellar objects.

\subsubsection{Protonated formaldehyde}
\label{sec3.1.3}

\begin{figure}
\centering{
\includegraphics[width=17cm,keepaspectratio]{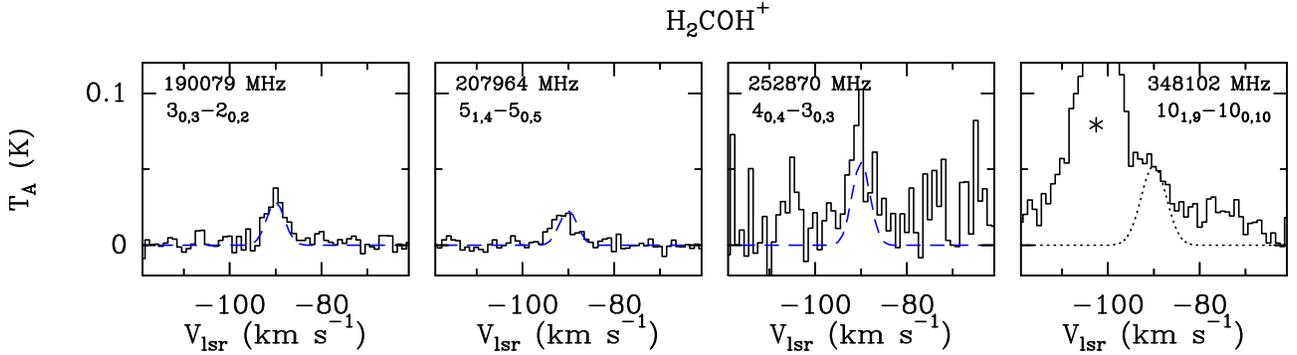}
\caption{Legend similar to Fig.~\ref{fig:fig2} but for the spectral lines of H$_2$COH$^+$ with their LTE models (dashed lines). In the fourth panel, the Gaussian fit depicts an H$_2$COH$^+$ transition tentatively identified at $\sim$~348102.330~MHz. The asterisk indicates a line likely blended with $^{34}$SO$_2$ at $\sim$~348117~MHz.}
\label{fig:fig6}}
\end{figure}

The search for H$_2$COH$^+$ was performed across the frequency range $\sim$~159--356 GHz. As a result, the identified lines of H$_2$COH$^+$ are exhibited in Fig.~\ref{fig:fig6} and were adjusted by means of Gaussian functions. The spectroscopic and fit parameters are listed in Table~\ref{tab:tab2}. In Fig.~\ref{fig:fig6}, the first three spectra were identified without blended emission at the rest frequencies $\sim$~190079.13, 207964.75 and 252870.34~MHz; on the fourth display, a tentative identification was made at $\sim$348102.33~MHz, the detection is unclear due to a spectral signature which dominates the whole spectrum, it is likely associated with $^{34}$SO$_2$ at $\sim$~348117~MHz (e.g. \citealt{Jewell1989}).  

In this work, the detection of H$_2$COH$^+$ is reported for the first time in a hot molecular core as G331. This cation was identified by \citet{ohi1996}, across the frequency range $\sim$~31--174~GHz, toward massive star-forming regions but not in cold and dark clouds.  \citet{bac2016} detected H$_2$COH$^+$, across the frequency range $\sim$~102--168~GHz, in a cold ($\sim$~10~K) source, the prestellar core L1689B. By means of a different technique of observation, \citet{Meier1993} detected H$_2$COH$^+$ in the coma of comet P/Halley through a Neutral Mass Spectrometer on the Giotto spacecraft. 

\begin{table}
\caption{Spectral line analysis of the molecular ions HCO$^+$, H$^{13}$CO$^+$, HC$^{18}$O$^+$, HC$^{17}$O$^+$ and H$_2$COH$^+$.}
\label{tab:tab2}
\centering
\begin{tabular}{lllcccccc}
\hline\hline 
Species and transitions & Frequency & Beamwidth & $E_u$ & $A_{ul}$ & Line area  &  $V_{\rm lsr}$ & Linewidth  & rms\\
 & [MHz]     & [\arcsec] &[K] & [10$^{-5}$s$^{-1}$] & [K km s$^{-1}$] &  [km s$^{-1}$] & [km s$^{-1}$] & [mK]    \\
\hline   
HCO$^+$ (2	--	1)$^a$	&	178375.056	&	34.98  	&	12.84	&	40.2	&	20	$\pm$	2	&	-91.56	$\pm$	0.02	&	6.97	$\pm$	0.07	& 120 \\
HCO$^+$ (3	--	2)$^a$ 	&	267557.626	&	23.32  	&	25.68	&	145	&	31	$\pm$	3	&	-91.21	$\pm$	0.03	&	9.2	$\pm$	0.1	& 133\\
\hline
H$^{13}$CO$^+$ (2	--	1)	&	173506.700	&	35.96  	&	12.49	&	37	&	6.41	$\pm$	0.02	&	-89.631	$\pm$	0.006	&	4.93	$\pm$	0.01	& 13 \\
H$^{13}$CO$^+$ (3	--	2)	&	260255.339	&	23.97  	&	24.98	&	134	&	7.05	$\pm$	0.05	&	-89.94	$\pm$	0.02	&	5.14	$\pm$	0.04	& 53\\
H$^{13}$CO$^+$ (4	--	3)	&	346998.344	&	17.98  	&	41.63	&	329	&	6.25	$\pm$	0.02	&	-90.205	$\pm$	0.007	&	5.19	$\pm$	0.02	& 22\\
\hline
HC$^{18}$O$^+$ (2	--	1)	&	170322.626	&	36.63  	&	12.26	&	35	&	1.05	$\pm$	0.01	&	-89.69	$\pm$	0.03	&	4.37	$\pm$	0.07	& 11\\
HC$^{18}$O$^+$ (3	--	2)	&	255479.389	&	24.42  	&	24.52	&	127	&	1.14	$\pm$	0.04	&	-89.86	$\pm$	0.08	&	4.5	$\pm$	0.2	& 44\\
\hline
HC$^{17}$O$^+$ (2	--	1)	&	174113.169	&	35.83  	&	12.53	&	37.4	&	0.31	$\pm$	0.01	&	-89.5	$\pm$	0.1	&	5.1	$\pm$	0.3	& 11\\
\hline
H$_2$COH$^+$	(3	0	3	--	2	0	2)	&	190079.131	&	32.82  	&	18.26	&	6.59	&	0.16	$\pm$	0.01	&	-89.9	$\pm$	0.2	&	4.5	$\pm$	0.5	&  12\\
H$_2$COH$^+$	(5	1	4	--	5	0	5)	&	207964.754	&	30.00  	&	55.51	&	14.7	&	0.12	$\pm$	0.01	&	-91.6	$\pm$	0.2	&	5.6	$\pm$	0.8	&   17\\
H$_2$COH$^+$	(4	0	4	--	3	0	3)	&	252870.339	&	24.67  	&	30.39	&	16.1	&	0.23	$\pm$	0.05	&	-90.7	$\pm$	0.3	&	3.2	$\pm$	0.9	&  44\\
H$_2$COH$^+$	(10	1	9	--	10	0	10)$^b$	&	348102.330	&	17.92  	&	181.67	&	48.2	&	0.4	$\pm$	0.1	&	-91.6	$\pm$	0.4	&	7	$\pm$	2	& 30\\
\hline
\end{tabular}
a)The line fit parameters of the HCO$^+$ lines were obtained using Lorentzian functions.
b)Line likely blended  with $^{34}$SO$_2$.
\end{table}

\begin{figure}
\centering{
\includegraphics[width=5.5cm,keepaspectratio]{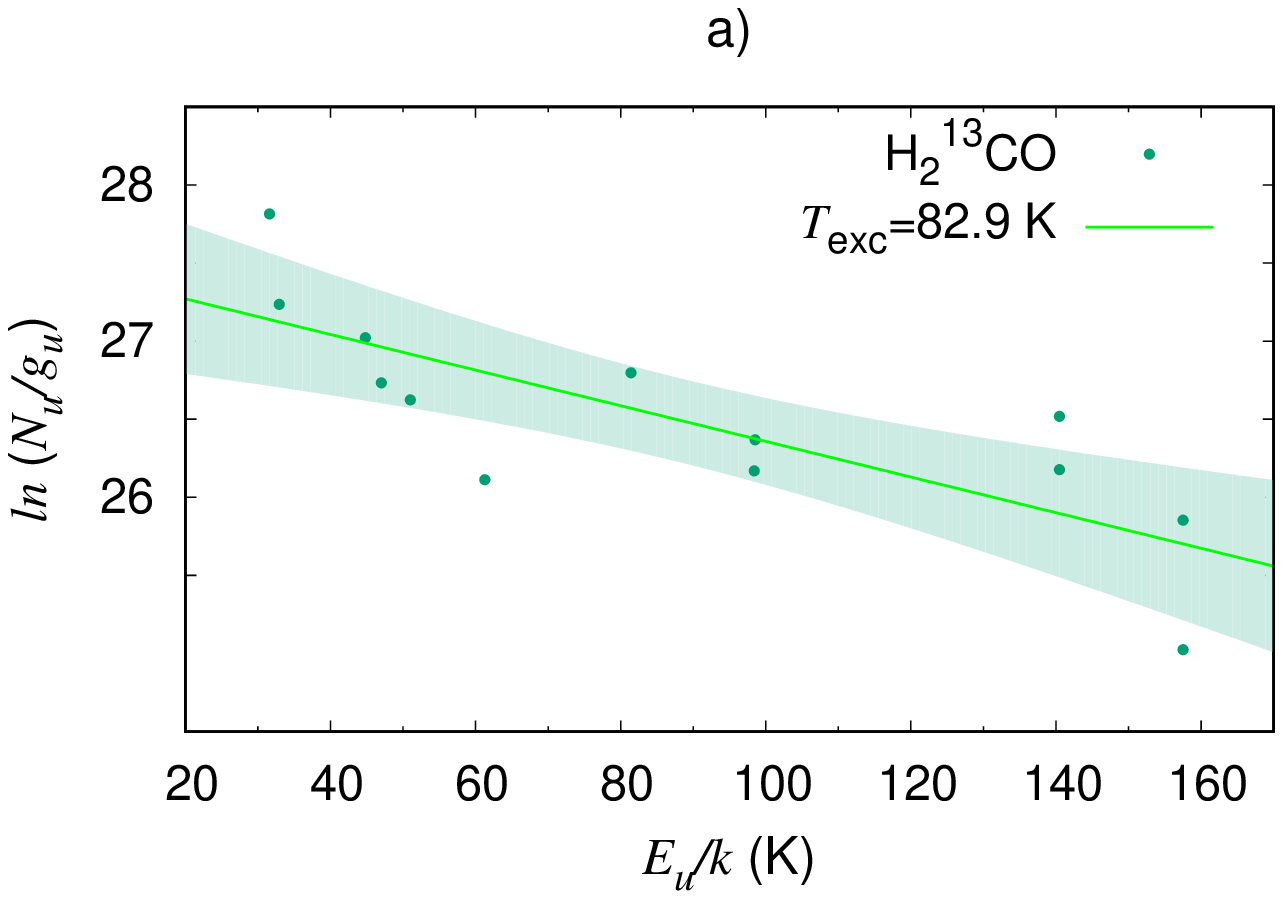}
\includegraphics[width=5.5cm,keepaspectratio]{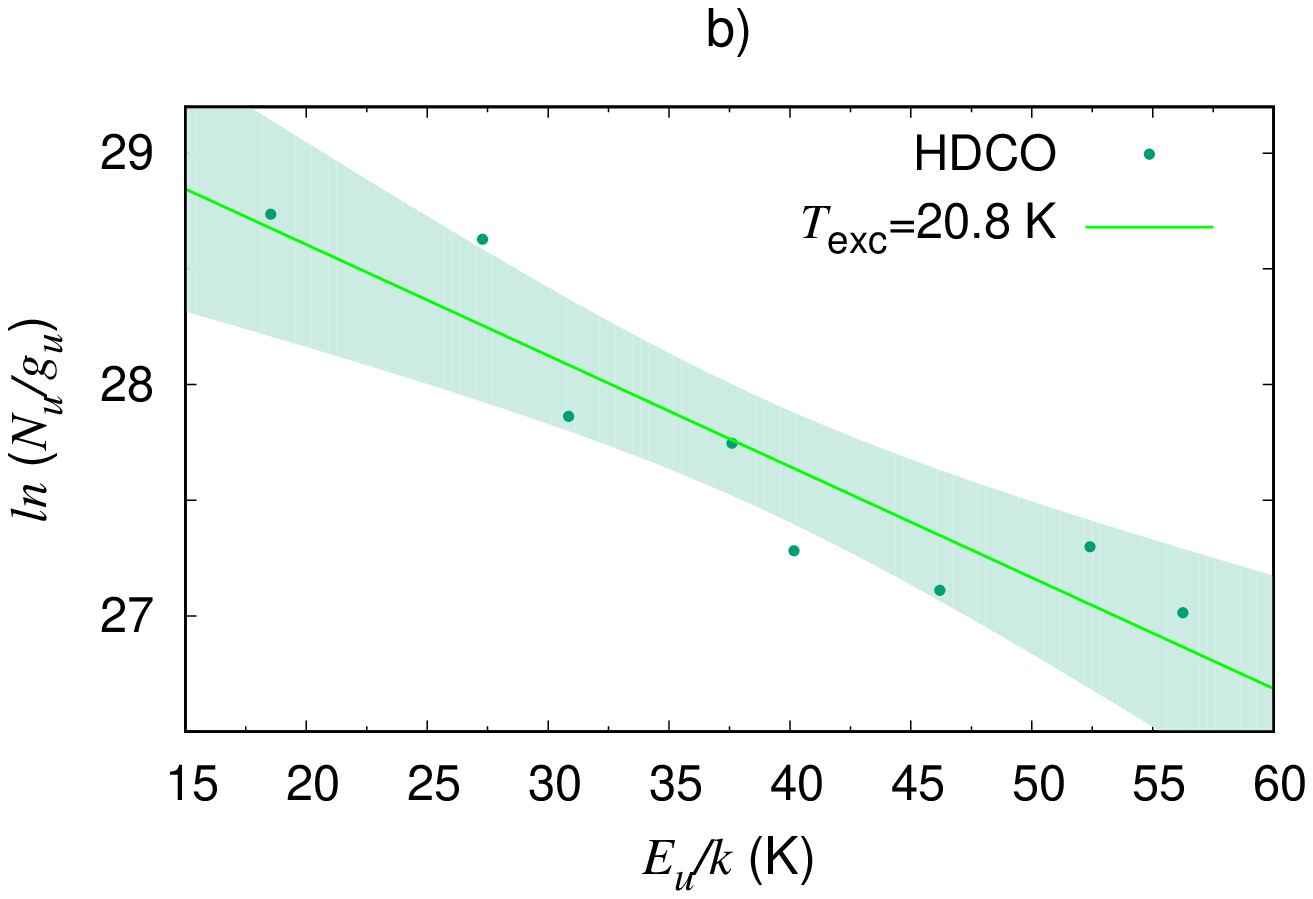}
\includegraphics[width=5.5cm,keepaspectratio]{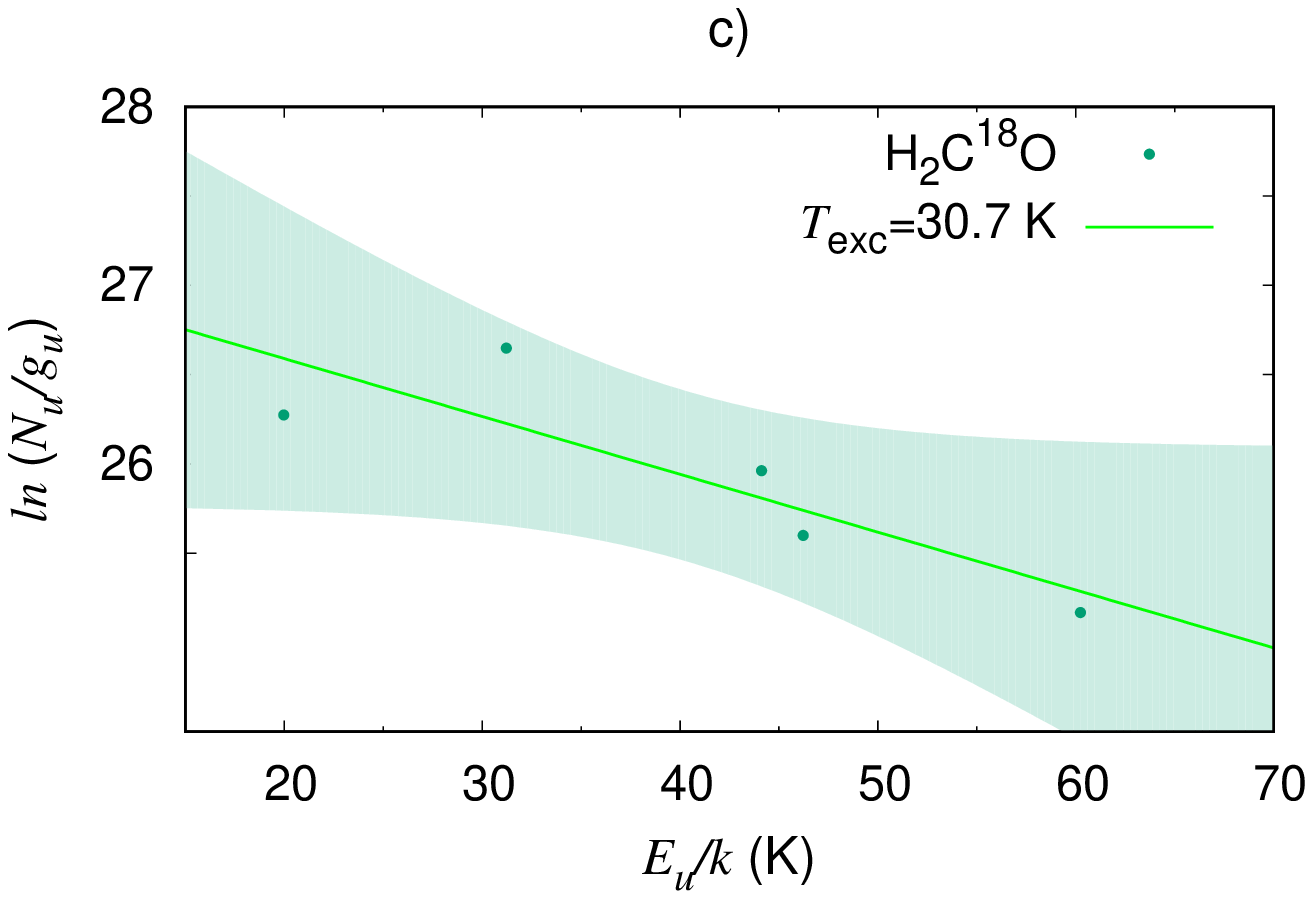}\\
\includegraphics[width=5.5cm,keepaspectratio]{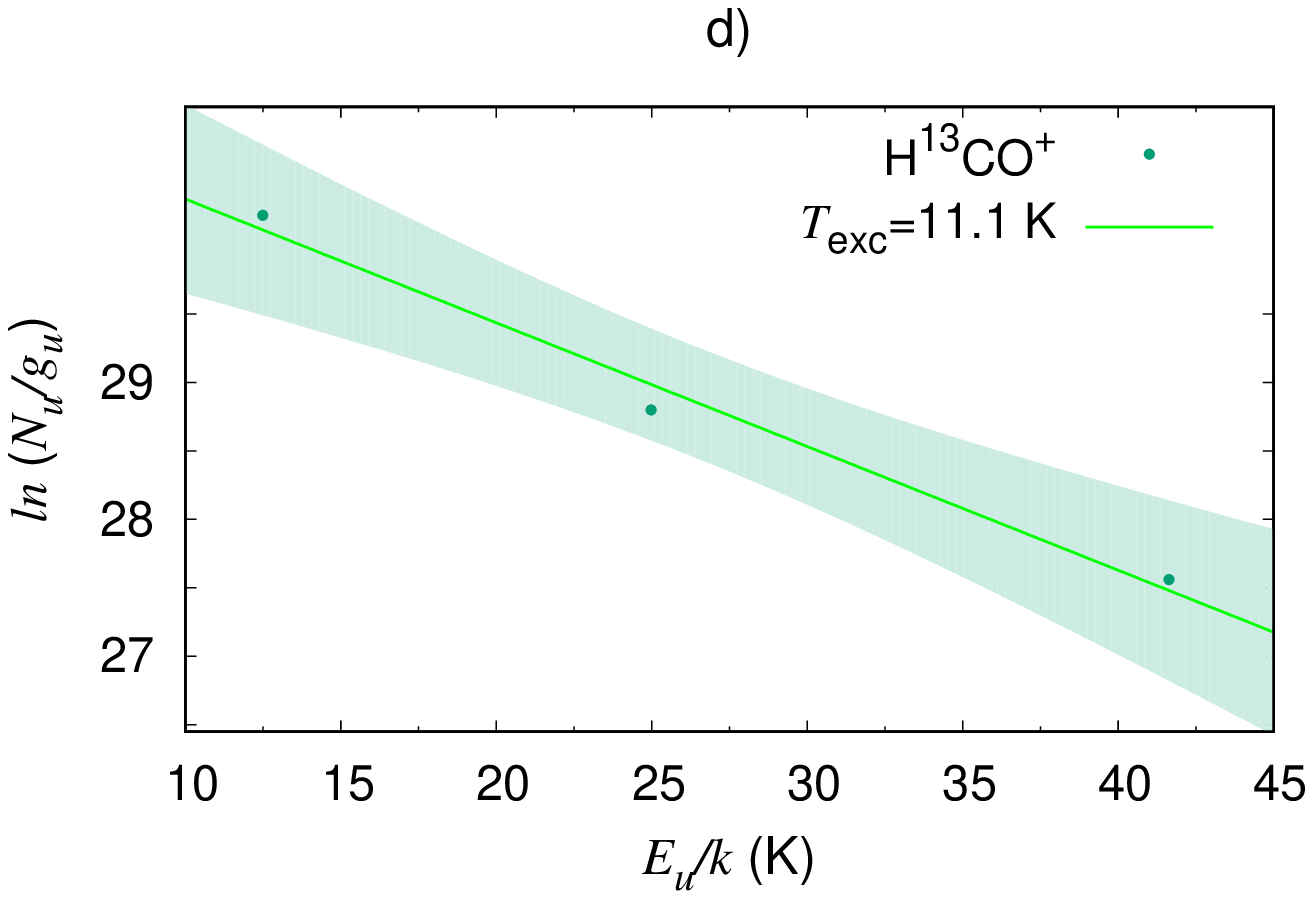}
\includegraphics[width=5.5cm,keepaspectratio]{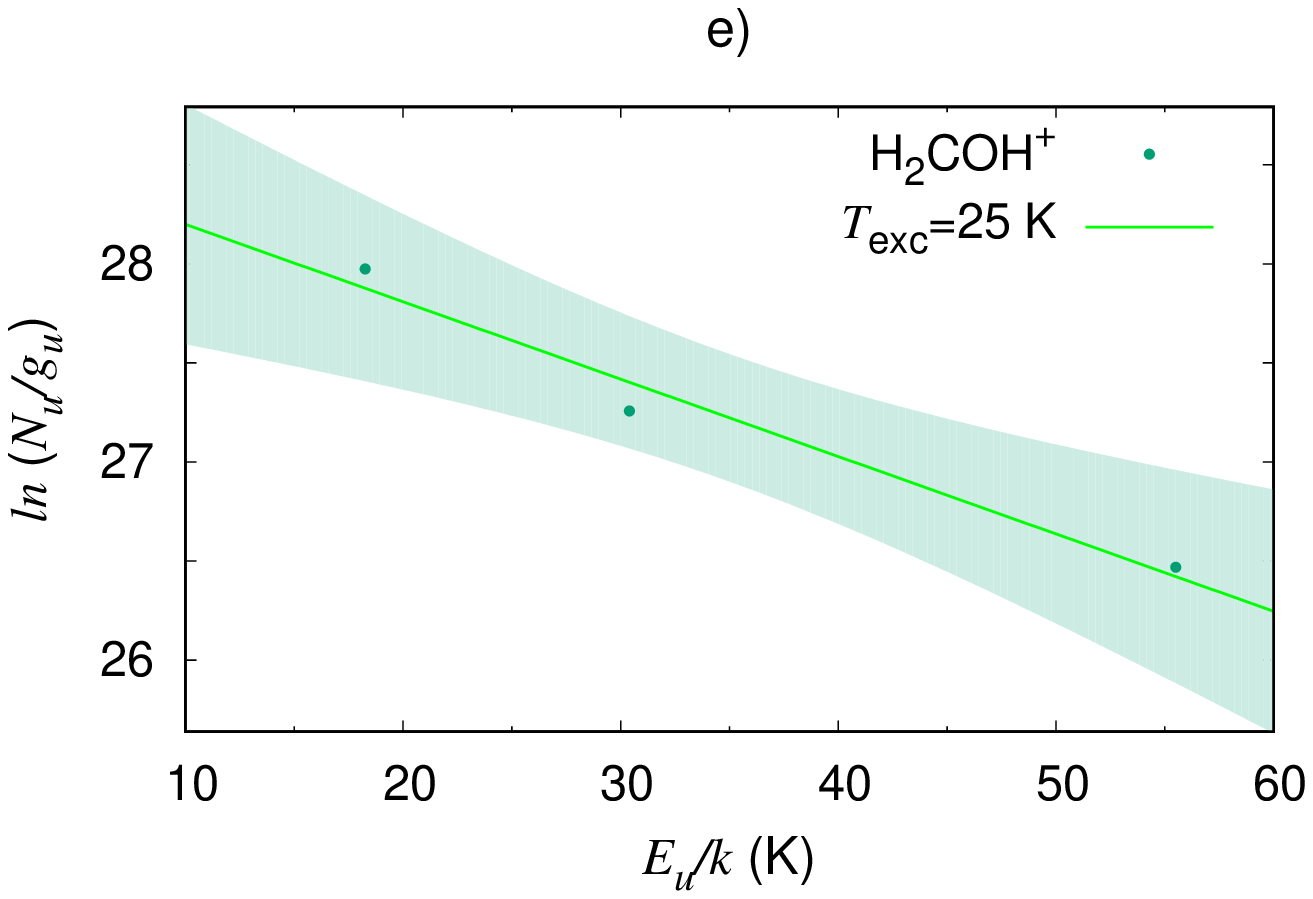}
\caption{Population diagrams of the formaldehyde isotopologues a) H$_2^{13}$CO, b) HDCO and c) H$_2$C$^{18}$O; and of the molecular ions d) H$^{13}$CO$^+$ and e) H$_2$COH$^+$. The solid-line and shaded areas represent the linear regression fits and their (95\%) confidence bands for each species.}
\label{fig:fig7}}
\end{figure}

\subsection{Physical conditions} \label{sec3.2}

The temperature and the column densities of the molecular species were estimated using LTE and Non-LTE conditions.
LTE methods were applied to analyze the physical conditions of all the detected isotopologues of formaldehyde, the formyl cation and the protonated formaldehyde (H$_2$COH$^+$). Non-LTE hypotheses were  considered to infer the physical conditions of the $^{12}$C isotopologues, H$_2$CO and HCO$^+$. All the analyses were performed under the general assumption of a source size of 5\arcsec. In the particular case of HCO$^+$, a second solution adopting 15\arcsec  was tested too.

When the LTE approximation is assumed, the population diagram method was used to estimate the excitation conditions of the molecular species, i.e., the total column densities ($N$) and excitation temperatures ($T_{\rm exc}$) by means of 

\begin{equation}
\ln \frac{N_u}{g_u} = \ln \frac{N}{Q} - \frac{E_u}{k T_{\rm exc}}, 
\label{eq1}
\end{equation}

\noindent where  $N_u$, $g_u$, $E_u$ and $Q$ are the column density, the degeneracy of the upper state, the energy of the upper level involved in the transition and the partition function, respectively. In Table~\ref{tab:tab1}, the upper state degeneracy $g_u=(2J_u+1)g_{ns}^{(u)}$ was given for each detected transition, where $J_u$ and $g_{ns}^{(u)}$ are the rotational angular momentum and the nuclear spin statistical weight of the upper state~\citep{BJbook}.
 The optical depth ($\tau$) has been inferred using the population diagram method. For a given transition, this can be estimated using the following expression

\begin{equation}
\tau = \frac{c^3 A_{ul} N_u}{8 \pi \nu^3 \Delta v \sqrt{\pi}/2\sqrt{\ln 2}} [\exp(h\nu/kT_{\rm exc})-1],
\end{equation}

\noindent where $\Delta v$ and $A_{ul}$ represent the spectral linewidth (km s$^{-1}$) and the Einstein coefficients (s$^{-1}$), respectively  \citep{Goldsmith1999,Mangum2015,Vastel2015}.

\subsubsection{Formaldehyde isotopologues: LTE analyses}

The population diagrams of the  $^{13}$C, D and $^{18}$O isotopologues of formaldehyde have been obtained with updated values of the partition function (see \ref{appendix}) and Eq.~(\ref{eq1})
can be corrected taking into account the optical depth ($\tau$) applied when the gas emission tends to be optically thick, and beam dilution factors, which allows us to  constrain the size of the emitting region with respect to the antenna beam \citep{Goldsmith1999}. The population diagrams of H$_2^{13}$CO, HDCO and H$_2$C$^{18}$O are exhibited in Fig.~\ref{fig:fig7} (panels a, b and c, respectively) and were obtained after applying a beam dilution correction for an emitting source assumed to be 5\arcsec \ and observed with an antenna beam ranged from $\sim$~17 to 39\arcsec.

\underline{H$_2^{13}$CO:} Its population diagram was plotted using the 13 detected lines corresponding to transitions in the $E_u \thickapprox 31-158$~K range (Fig.~\ref{fig:fig7}a). The linear fit provided the result $T_{\rm exc}=82.9 \pm 21.8$~K and $N=(8.0\pm 2.0) \times 10^{14}$~cm$^{-2}$ ($\chi^2_{\rm red}$=1.53). From this result, synthetic spectra were computed and compared with the observed lines in Fig.~\ref{fig:fig2}. The values of the optical depth were estimated $\tau \lesssim 1$ for all the spectra, except for the partially blended lines at $\sim$~284117.45 and 284120.62~MHz, in agreement with the LTE~formalism. 

In comparison with previous works, \citet{Schoier2002} carried out radiative analyses of H$_2^{13}$CO based on transitions in the $E_u \thickapprox 21-99$~K range toward the low-mass protostellar object IRAS 16293-2422. They reported an excitation temperature around 90~K. 

\underline{HDCO:} Based on its 8 spectral lines, detected in the range $E_u \thickapprox$ 18--56~K, the population diagram is displayed in Fig.~\ref{fig:fig7}(b). From the linear fit, it was obtained $T_{\rm exc}$ = 20.8 $\pm$ 3.5~K and $N$ = (2.5 $\pm$ 0.8) $\times$ 10$^{14}$~cm$^{-2}$ ($\chi^2_{\rm red}$=0.84). The observed spectral lines and the simulated ones obtained from this result are exhibited in Fig.~\ref{fig:fig2}, the values of the optical depth were estimated $\tau \lesssim 1$.

\citet{Neill2013} performed a study about deuterated molecules in Orion KL. They discussed for H$_2^{13}$CO and HDCO different LTE scenarios with temperatures of $\sim$~40, 63, and 67~K as well as particular aspects about the deprotonation of H$_2$COH$^+$ and HDCOH$^+$. \citet{Bianchi2017} analyzed several deuterated species toward the Class I protostar SVS13-A. From the population diagrams of H$_2^{13}$CO, HDCO and D$_2$CO, with source sizes of $\sim$~10\arcsec, they derived temperatures of $\sim$~23, 15, and 28~K, respectively.

\underline{D$_2$CO:} LTE calculations were carried out to estimate an upper limit on the D$_2$CO emission. Taking into account the excitation conditions of HDCO, the D$_2$CO column density was computed assuming that $N$(D$_2$CO) $< N$(HDCO)$\thickapprox$2.5$\times$10$^{14}$~cm$^{-2}$ and $T_{\rm exc}\thickapprox$ 21~K. The best solution gave $N$(D$_2$CO) $\thickapprox$ 1.3 $\times$ 10$^{14}$~cm$^{-2}$ ($\chi
^2_{\rm red}$=0.8) providing a column density ratio of D$_2$CO/HDCO $\thickapprox$ 0.5.\\

D$_2$CO has been observed in only few sources. In Orion~KL, \citet{Neill2013} established an upper limit on the D$_2$CO abundance of [D$_2$CO]/[HDCO]$\leq$0.1.  \citet{Turner1990} detected three transitions of D$_2$CO in Orion KL reporting [D$_2$CO]/[HDCO]=0.02. In IRAS 16293-2422, a source with several studies on molecular deuteration, \cite{Ceccarelli1998} found [D$_2$CO]/[HDCO]$\leq$0.5. 

\underline{H$_2$C$^{18}$O:} The population diagram of this isotopologue was obtained from 5 detected transition lines which covered the $E_u \thickapprox$ 19--60~K range. The linear fit provided $T_{\rm exc}$ = 30.7 $\pm$ 10.3~K and $N$ = (6.8 $\pm$ 3.2) $\times$ 10$^{13}$~cm$^{-2}$ ($\chi
^2_{\rm red}$=1.23). From this solution, synthetic spectra were simulated and compared with the observed lines (Fig.~\ref{fig:fig2}). The spectral lines of H$_2$C$^{18}$O are optically thin, as expected according to its low abundance. It is worth mentioning  that the population diagrams of HDCO and H$_2$C$^{18}$O have similar $E_u$ ranges, between~$\sim~18-60$~K, different from  those of H$_2^{13}$CO within $E_u \sim$~30--160~K. The excitation temperatures of HDCO and H$_2$C$^{18}$O ($T_{\rm exc} \lesssim$ 30~K) are lower than the obtained one for  H$_2^{13}$CO ($T_{\rm exc} \thickapprox$ 83~K). On the one hand, such difference might come from the fact that more lines of H$_2^{13}$CO were observed than HDCO and H$_2$C$^{18}$O, providing a more extended population diagram. On the other hand, there might be a scenario in which the formaldehyde isotopologues trace different excitation conditions although, to assume this hypothesis would require further studies including, e.g., gas-grain calculations and chemical models.

\subsubsection{H$_2$CO: Non-LTE calculations}

The physical conditions of H$_2$CO were calculated using the non-LTE code RADEX,  which provides an alternative approach to the population diagram based on the assumption of optically thin emission lines. Within this approach, the optical depth effects are treated with an escape probability method  
assuming an isothermal and homogeneous medium without large-scale velocity fields~\citep{van2007}. The non-LTE calculations  were performed using available excitation rates between o- and p-H$_2$CO and o- and p-H$_2$. \citet{Wiesenfeld2013} reported calculations for the first 81 rotational levels of formaldehyde considering temperatures between $\sim$~10 and 300~K. In an earlier work, \citet{Troscompt2009} also discussed the rotational excitation of H$_2$CO by H$_2$. 

From the five lines of p-H$_2$CO, the best solution ($\chi^2_{\rm red}$=4.5)  provided $T_{\rm kin} = 87.4 \pm 0.6$~K and $N= (6.2 \pm 0.3) \times 10^{15}$~cm$^{-2}$. From the four lines of o-H$_2$CO, it was obtained $T_{\rm kin} = 95.1 \pm 0.3$~K and $N= (9 \pm 3) \times 10^{15}$~cm$^{-2}$ ($\chi^2_{\rm red}$=3.5). These results provide a ortho-to-para ratio of H$_2$CO of~$\sim$~1.4. The H$_2$ density was also estimated from the calculations $n({\rm H_2}) \simeq 4  \times 10^7$~cm$^{-3}$ adopting a source size of 5\arcsec. The density of molecular hydrogen was computed using conventional ortho-to-para ratios, whose value can fluctuate between $\sim$0.1 and 3, contingent on the chemistry, the physical properties and the evolutionary stage of sources. In prestellar cores, such a range was discussed from 3 to values even smaller than 0.001 (e.g. \citealt{Osterbrock1962,lac1994,Pagani2013}).

Synthetic lines were generated from the non-LTE approximations and compared with the observations in  Fig.~\ref{fig:fig2}(a). Concerning the optical depth of the p-H$_2$CO transitions, the highest and lowest values were~$\sim$~4.5 and 0.9 for the lines at $\sim$~290623.40~MHz and 218760.06~MHz, respectively. Similarly, for the o-H$_2$CO transitions, the highest and lowest optical depths were $\sim$~5.2 and $\sim$~0.7 for the transitions at $\sim$~351768.64~MHz and 291380~MHz, respectively.

A comparison of our results with other works is useful to show they are in accordance. \citet{Schoier2002} analysed the physical conditions of  H$_2$CO in IRAS 16293-2422. They estimated temperatures around 90~K and ortho-to-para ratios of H$_2$CO around~0.9. \citet{Guzman2013} used non-LTE models of  o- and p-H$_2$CO based on observations with the IRAM-30m toward the Horsehead photo-dissociation region (PDR) to estimate ortho-to-para ratios around 3 at the dense core and 2 in the PDR. In a pioneering work, \citet{Mangum1990} stressed the importance of H$_2$CO as a key tracer to estimate density, temperature and molecular abundances. They performed LTE and non-LTE calculations obtaining $n$(H$_2$)=(0.5--1)$\times$10$^7$~cm$^{-3}$ and $T_{\rm kin} \thickapprox$100~K in Orion KL. In this work we detected various lines of formaldehyde that were identified in Orion KL by \citet{Mangum1990}. 

\subsubsection{Formyl cation isotopologues: LTE analyses} \label{sec3.2.3}

The population diagram obtained for  H$^{13}$CO$^+$ using the three detected lines is exhibited in Fig.~\ref{fig:fig7}(d). Assuming that the emitting region is 5\arcsec, the excitation temperature and the column density were estimated,  $T_{\rm exc}$ = 11 $\pm$ 1~K and $N$ = (2.1 $\pm$ 0.7) $\times$ 10$^{14}$~cm$^{-2}$ ($\chi^2_{\rm red}$=1.30), respectively. In the case of HC$^{18}$O$^+$ and HC$^{17}$O$^+$, no population diagram could be displayed because of the lack of enough detected lines. Nevertheless, LTE calculations were performed to estimate their column densities. Similar methodologies were discussed in \citet{Schoier2002}, who performed LTE and non-LTE calculations for HCO$^+$ and its isotopologues in IRAS 16293-2422. 

In this work the LTE calculations of HC$^{18}$O$^+$ and HC$^{17}$O$^+$ were performed fixing the excitation temperature to the value obtained for H$^{13}$CO$^+$ and considering the column densities as free parameters with the following condition $N$(HC$^{17}$O$^+$) $< N$(HC$^{18}$O$^+$) $< N$(H$^{13}$CO$^+$). The estimates of the column densities were $N$(HC$^{18}$O$^+$) $\thickapprox$ 1.8 $\times$10$^{13}$~cm$^{-2}$ ($\chi
^2_{\rm red}$=1.5) and $N$(HC$^{17}$O$^+$) $\thickapprox$ 1.4 $\times$10$^{13}$ cm$^{-2}$ ($\chi
^2_{\rm red}$=0.3). Synthetic spectra were included in the panels of Fig.~\ref{fig:fig5}. 

The low temperature estimated from the  H$^{13}$CO$^+$ lines suggests that its emission is likely  associated  with a cold and expanded region. In addition, we also obtained results assuming an extended source size of 15\arcsec. Thus, the  H$^{13}$CO$^+$ population diagram gave the values $T_{\rm exc}$ = 12 $\pm$ 2~K and $N$ = (2.6 $\pm$ 0.8) $\times$ 10$^{13}$~cm$^{-2}$,  and for the $^{18}$O and $^{17}$O isotopologues $N$(HC$^{18}$O$^+$) $\thickapprox$ 4.2 $\times$10$^{12}$ cm$^{-2}$ and $N$(HC$^{17}$O$^+$) $\thickapprox$ 1.5 $\times$10$^{12}$ cm$^{-2}$. 

\subsubsection{HCO$^+$: Non-LTE calculations} \label{sec3.2.4}

The main isotopologue HCO$^+$ was observed through two intense and optically thick lines (Fig.~\ref{fig:fig4}), they were analyzed via  non-LTE calculations using collisional excitation rates.
The two transitions 2--1 and 3--2 were computed using RADEX and the rate coefficients available in the LAMDA database~\citep{Flower1999} assuming $n$(H$_2$)=(0.1--5)$\times$10$^7$~cm$^{-3}$, $T <$~30~K and 5\arcsec. The column density resulted in $N$(HCO$^+$) $\thickapprox$ 3 $\times$ 10$^{15}$ cm$^{-2}$ ($\chi^2_{\rm red}$=6.5). In the literature, line analysis of HCO$^+$ has been reported with high reduced values of $\chi^2$ \citep{Schoier2002}. The result reported in this work should be taken as a rough approximation.

\subsubsection{Protonated formaldehyde: LTE analysis}

The 3 unblended lines of H$_2$COH$^+$ allowed us to obtain the population diagram shown in Fig.~\ref{fig:fig7}(e) and the estimates $T_{\rm exc}$ = 25 $\pm$ 4~K and $N$ = (1.4 $\pm$ 0.3) $\times$ 10$^{14}$~cm$^{-2}$ ($\chi^2_{\rm red}$=0.41) assuming a source size of 5\arcsec. Fig.~\ref{fig:fig6} exhibits a comparison between the simulated and the observed lines with the exception of the spectrum affected by contaminant emission.

In comparison with other works, \citet{ohi1996} reported $T_{\rm exc}=60-110$~K and  $N$(H$_2$COH$^+$)$\thickapprox$10$^{12}$--10$^{14}$ cm$^{-2}$ in surveys toward Sgr~B2, Orion~KL and W51. In W51, since they could not detect a sufficient number of lines of H$_2$COH$^+$, they used HCO$^+$ to infer the physical conditions of H$_2$COH$^+$ taking different excitation temperatures. In the ultra cold ($T \sim$10 K) source L1689B, \citet{bac2016} estimated column densities between 3 $\times$ 10$^{11}$~cm$^{-2}$ and 1 $\times$ 10$^{12}$~cm$^{-2}$. They discussed that the H$_2$COH$^+$ formation can occur via:

\begin{equation}\label{Eq:3}
\text{H$_2$CO\ +\ x$^+$ $\longrightarrow$ H$_2$COH$^+$ + H$_2$},
\end{equation}

\noindent where x$^+$ represents a proton donor such as H$_3^+$. Furthermore, in the context of Eq.~\ref{Eq:1}, the specific proton donor is HCO$^+$ obtaining CO  as a by-product instead of H$_2$. They predicted the abundance ratios [H$_2$COH$^+] \thickapprox$ 0.007[H$_2$CO], when H$_3^+$ is the proton donor, and [H$_2$COH$^+] \thickapprox$ 0.003 [H$_2$CO], when it is HCO$^+$. 

In summary, we present in Table~\ref{tab:tab3} the results obtained from the LTE and non-LTE analyses of the isotopologues of formaldehyde and the formyl. The results suggest that these species might trace different gas components. From the estimated column densities of the formaldehyde isotopologues, it was obtained the abundance ratios H$_2$CO:H$_2^{13}$CO:HDCO:H$_2$C$^{18}$O$\thickapprox$223:12:4:1. In addition, Table~\ref{tab:tab3} also presents the results obtained from the population diagram of H$_2$COH$^+$. 

\begin{table}
\centering
\caption{Summary of the LTE and non-LTE analyses of the isotopologues of formaldehyde and the formyl cation, and the protonated formaldehyde. The temperatures and column densities come from the results obtained at a source size of 5\arcsec.}
\label{tab:tab3}
\centering
\begin{tabular}{l l l l l}
\hline\hline
Species  & \# Analysed & Method     & $N$ [cm$^{-2}$] & $T$ [K]\\
          & lines    &            & $\thickapprox$        & $\thickapprox$ \\
\hline
H$_2$CO	&	9	&	non-LTE$^a$	&	1.52 $\times$ 10$^{16}$	&	91	\\
H$_2^{13}$CO	&	13	&	LTE$^b$	&	8.0 $\times$ 10$^{14}$	&	83	\\
HDCO	&	8	&	LTE$^b$	&	2.5 $\times$ 10$^{14}$	&	21	\\
H$_2$C$^{18}$O	&	5	&	LTE$^b$	&	6.8  $\times$ 10$^{13}$	&	31	\\
D$_2$CO	&	2	&	LTE$^c$	&	$<$1.3  $\times$ 10$^{14}$	&	21	\\
\hline
HCO$^+$	&	2	&	non-LTE	&	3$\times$10$^{15}$	&	$<$30			\\
H$^{13}$CO$^+$	&	3	&	LTE$^b$	&	2.1$\times$10$^{14}$	&	11			\\
HC$^{18}$O$^+$	&	2	&	LTE$^d$	&	1.8$\times$10$^{13}$	&	11			\\
HC$^{17}$O$^+$	&	1	&	LTE$^d$	&	1.4$\times$10$^{13}$	&	11			\\
\hline
H$_2$COH$^+$	&	3	&	LTE$^b$ & 1.4 $\times$ 10$^{14}$& 25\\
\hline
\end{tabular}\\
a) From the o-H$_2$CO and p-H$_2$CO results, the column density and temperature are the sum and mean value, respectively. b) Obtained from the population diagram analysis. c) Upper limit based on the HDCO analysis. d) Obtained from the LTE calculation. 
\end{table}

\section{Discussion}\label{sec4}

\subsection{Fractional abundances}

The fractional abundances ([$X$]) were estimated with respect to molecular hydrogen using [$X$]=$N_i$/$N_{\text{H$_2$}}$, where $N_i$ and $N_{\text{H$_2$}}$ represent the column density of the  species $i$ and H$_2$, respectively. The H$_2$ column density was indirectly inferred from H$^{13}$CO$^+$ by means of the ratio H$^{13}$CO$^+$/H$_2$  = 3.3 $\times$ 10$^{-11}$ \citep{bla1987,Sanchez2013,mer2013b}. Using that ratio and the results obtained from the population diagrams of H$^{13}$CO$^+$, considering source sizes of 5 and 15\arcsec,  an interval of H$_2$ column densities was  estimated as $N_{\text{H$_2$}}= (0.8-6) \times
10^{24}$~cm$^{-2}$. In the literature, similar H$_2$ column densities have been discussed in the context of high-mass star-forming regions (e.g. \citealt{Yu2018,Motte2018}). In addition, values of the order of $N_{\text{H$_2$}} \thickapprox 10^{24}$~cm$^{-2}$ have been used in Orion KL (e.g. \citealt{Crockett2014}). In
Table~\ref{tab:tab4}, we summarize the abundances of the main isotopologues H$_2$CO, HCO$^+$ and H$_2$COH$^+$ obtained with respect to the estimated $N_{\text{H$_2$}}$ interval and are compared with values reported for the sources NGC 7129 FIRS 2, Orion KL, Sgr B2 and W51, which are known sources for exhibiting a rich chemistry in simple and complex organic molecules \citep{Blake1986,ohi1996,fuen2014,Crockett2014}. 

\begin{table}
\centering
\caption{Fractional abundances of  H$_2$CO, HCO$^+$ and H$_2$COH$^+$ in G331 in comparison with values reported in other sources.}
\label{tab:tab4}
\centering
\begin{tabular}{l l l l}
\hline\hline
Species    & \multicolumn{3}{c}{Fractional abundances}\\
    \cline{2-4}
  &  This work   & Other works  & Ref.  \\
\hline
H$_2$CO    & (0.2--2) $\times$10$^{-8}$  & (2--8) $\times$ 10$^{-8}$    &(1)         \\
HCO$^+$    & (0.5--4) $\times$10$^{-9}$
         &  2.3 $\times$ 10$^{-9}$    &(2)          \\
H$_2$COH$^+$  & (0.2--2) $\times$10$^{-10}$
         & (0.01--1) $\times$10$^{-9}$ &(3)       \\  
\hline
\end{tabular}\\
(1) NGC 7129 FIRS 2 \citep{fuen2014}. (2) OMC-1, extended ridge \citep{bla1987}. (3) Sgr B2, Orion KL, and W51 \citep{ohi1996}.
\end{table}

\subsection{Chemical modelling} \label{sec4.2}

In the gas phase, there are several ion-molecule reactions that can explain the H$_2$COH$^+$ formation. In Eq.~(\ref{Eq:1}), one of the most important mechanism involving the chemical species H$_2$CO and HCO$^+$ is described \citep{Tanner1979,Woon2009}. Concerning surface reactions,  \citet{Song2017} performed calculations about the hydrogenation of H$_2$CO in amorphous solid water surfaces and found some implications about the protonation of CH$_3$O isomers. H$_2$COH$^+$ is a major product from the ionization and fragmentation of CH$_3$OH, CH$_3$CH$_2$OH and CH$_3$OCH$_3$ \citep{Mosley2012}.

We observed that the reaction between H$_2$CO and HCO$^+$ is one of the most important channels to produce H$_2$COH$^+$, as well as the general scheme described in Eq.~\ref{Eq:3}. In addition, it is observed that other cations can also react with H$_2$CO to produce H$_2$COH$^+$, for instance:

\begin{eqnarray} 
\text{H$_2$CO + H$_3^+$} &\longrightarrow& \text{H$_2$COH$^+$ + H$_2$,}    \\ 
\text{H$_2$CO + H$_3$O$^+$} &\longrightarrow& \text{H$_2$COH$^+$ + H$_2$O,} \\ \label{Eq:prod}
\text{H$_2$CO + N$_2$H$^+$} &\longrightarrow& \text{H$_2$COH$^+$ + N$_2$.}
\end{eqnarray}

\noindent and the major reactions of destruction are

\begin{eqnarray}\label{Eq:7} 
\text{HCO$^+$ + e$^-$} &\longrightarrow&\text{CO + H,} \\ \label{Eq:8} 
\text{H$_2$COH$^+$ + e$^-$} &\longrightarrow&\text{HCO + 2H,} \\ \label{Eq:9} \text{H$_2$COH$^+$ + e$^-$} &\longrightarrow& \text{CO + H + H$_2$.}
\end{eqnarray}

\begin{figure}
\centering{
\includegraphics[width=8.5cm,keepaspectratio]{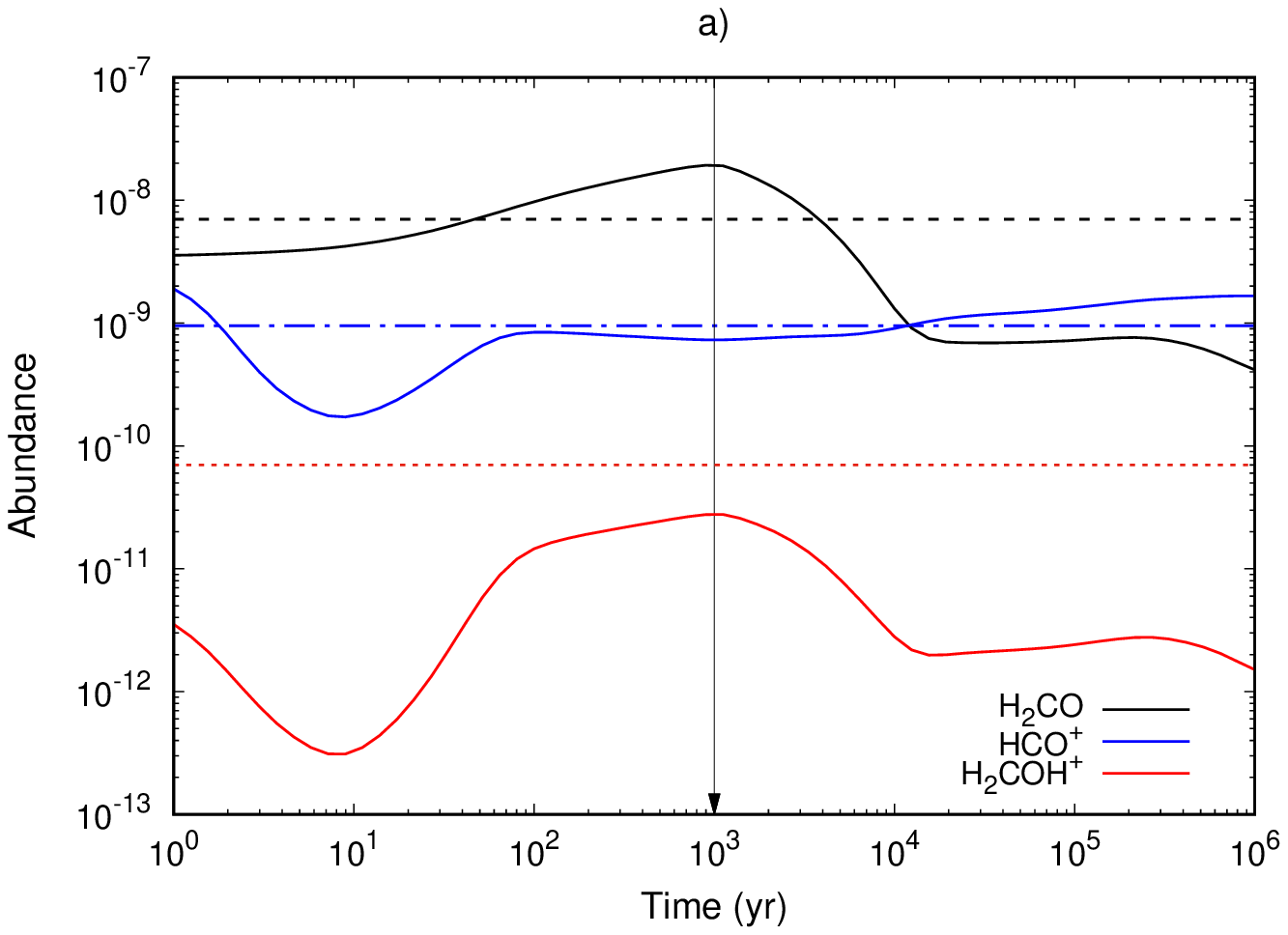}
\includegraphics[width=8.5cm,keepaspectratio]{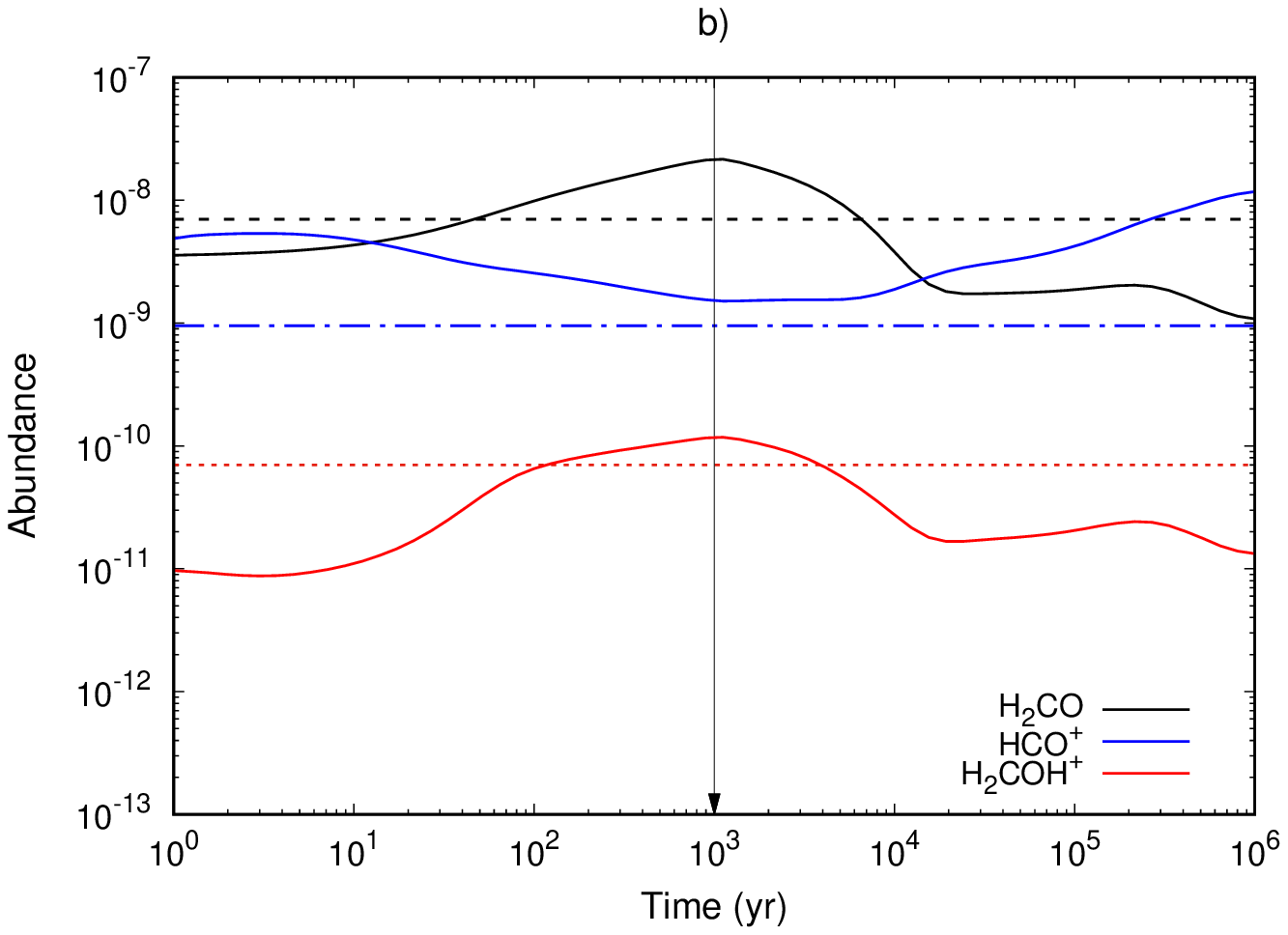}
\caption{Predicted abundances of H$_2$CO, HCO$^+$ and H$_2$COH$^+$ as a function of time, represented by the black, blue and red solid lines, respectively, obtained from a model with $T$=30~K and $n_{\text{H}_2}$=1 $\times$ 10$^6$~cm$^{-3}$. 
Panel a) represents a model with the whole chemical network whereas panel b) does not include the major destruction channels of HCO$^+$ and H$_2$COH$^+$. The dashed, dash-dotted and dotted  horizontal lines indicate the mean fractional abundances observed in G331 for H$_2$CO, HCO$^+$ and H$_2$COH$^+$, respectively. The vertical arrow depicts the  abundance peak of H$_2$CO at $t\thickapprox 10^3$~yr.}
\label{fig:fig8}}
\end{figure}

In this work, we performed a gas-grain model to study the chemistry of H$_2$CO, HCO$^+$ and H$_2$COH$^+$ (\S~\ref{sec2.1}). As a result, the evolution of the gas abundances were simulated for these species. Therefore, we adopted the initial chemical abundances of \citet{Vidal2018} (and references therein), but considering the physical parameters obtained in this work from the line observations of H$_2$CO, HCO$^+$ and H$_2$COH$^+$ in G331. The model was computed using  different values of gas temperature and density, assuming the ranges $T$=10--90~K and $n_{\text{H}_2}$ = (0.05--1) $\times$ 10$^7$ cm$^{-3}$, respectively, and using a visual extinction of $A_V=10$ mag and a cosmic ray ionization rate of $\zeta$=1.3$\times$10$^{-17}$~cm$^{-3}$.
The model that yielded the most accurate abundance prediction, consistent with the observed abundances within an order of magnitude, was obtained at $T$=30~K and $n_{\text{H}_2} = 1 \times$ 10$^6$ cm$^{-3}$. However, at temperatures and densities above these values, the model's chemistry become unpredictable. The result is exhibited in Fig.~\ref{fig:fig8}, where we marked the mean values of the fractional abundances of H$_2$CO, HCO$^+$ and  H$_2$COH$^+$ (given in Table~\ref{tab:tab4}) with horizontal lines (black dashed, blue dash-dotted and red dotted respectively). Here it is observed that H$_2$CO reaches its maximum abundance at $\sim$~10$^3$~yr (the time scale $\sim$~10$^3$~yr is pointed by a vertical line) in agreement with the results shown in Table~\ref{tab:tab4}. This timescale would be optimal for the formation of neutral species in  G331. Furthermore, at the time scale of $\sim$~10$^2$--10$^3$~yr, the predicted abundances of HCO$^+$ and H$_2$COH$^+$ are in agreement with  the observational mean values, in line with the general assumption of massive protostellar objects (i.e., young sources with an active chemistry with abundant molecular emission). These agreements make us develop the hypothesis that the lifetime of the source G331 might be $<$10$^4$~yr.

To explore the effects of the major reactions of destruction, we carry out a second simulation shown in Fig.~\ref{fig:fig8}. While Fig.~\ref{fig:fig8}(a) represents a model computed with the whole network of chemical reactions, Fig.~\ref{fig:fig8}(b) depicts a model that excludes the destruction reactions described in Eq.~\ref{Eq:7}, \ref{Eq:8}, and \ref{Eq:9}. In general,   the major destructive reactions  reduce noticeably the abundances of the molecular ions, but the abundance of neutral H$_2$CO remains relatively stable. In addition, in the models of Fig.~\ref{fig:fig8}(a) and Fig.~\ref{fig:fig8}(b), it is observed that H$_2$COH$^+$ follows the abundance curves of HCO$^+$ and H$_2$CO, respectively. 
In the case of Fig.~\ref{fig:fig8}(a),  HCO$^+$ is a major precursor of H$_2$COH$^+$. Thus, changes in the abundance of  HCO$^+$ affect  the production of H$_2$COH$^+$.  
In the case of Fig.~\ref{fig:fig8}(b), the increase of the abundance of HCO$^+$, which reacts with H$_2$CO to form H$_2$COH$^+$ (see Eq.~\ref{Eq:1}), make rise the abundance of H$_2$COH$^+$. 

The dissociative recombination of molecular ions with electrons is one of the most complex, destructive and least understood mechanism in the ISM \citep{Hamberg2007,Meier1993}. \citet{Hamberg2007} carried out experiments to measure the cross-sections and branching ratios of various protonated and deuterated molecular ions. \citet{Meier1993} detected H$_2$COH$^+$ in the coma of the comet P/Halley and performed an ion chemical model to estimate the H$_2$CO production from H$_2$COH$^+$. The chemical models presented in this work provide a preliminary framework for understanding the chemistry of molecular ions in cold regions of G331.

\subsection{Isotopic fractionation}

\underline{D/H ratios:} The ratio was calculated using the neutral isotopologues H$_2$CO and HDCO, while molecular ions HCO$^+$ and DCO$^+$ were not used due to the non detection of DCO$^+$ in G331. From the LTE and non-LTE analyses of HDCO and H$_2$CO, respectively, 
it is reported HDCO/H$_2$CO $\sim$~0.02 for a bulk gas with $T<$90~K and, from the D$_2$CO tentative detection, it was estimated D$_2$CO/H$_2$CO $\leq$ 0.009.  \citet{rob2007} estimated HDCO/H$_2$CO $\sim$~0.01, 0.03, and $<$0.002 in G34.26, G75.78 and G31.41, respectively. In addition, for the G34.26 and G75.78 sources, they also  reported D$_2$CO/H$_2$CO $<$~0.001 and $<$0.01, respectively, for gas components within a broad distribution of temperatures ($T<$~100~K).  

The HDCO detection demonstrated clear evidences of deuteration fractionation in G331. In previous studies of G331 conducted with APEX, deuterated species were searched but no conclusive results were obtained~\citep{men2018,duronea2019,Canelo2021,Santos2022}. 

\underline{$^{12}$C/$^{13}$C ratios:}  The H$_2$CO/H$_2^{13}$CO and HCO$^+$/H$^{13}$CO$^+$ ratios were estimated in G331 within $\sim$~14--20. \citet{Wilson1994} analyzed isotopic carbon ratios from CO and H$_2$CO  reporting values of $\sim$~20 and 77 for the Galactic center and Local ISM, respectively. \citet{Yan2019} estimated $^{12}$C/$^{13}$C ratios from formaldehyde.  From 112 observations, the transition  (1$_{1,0}$--1$_{1,1}$) of H$_2$CO was detected in 84 sources and from these 84, H$_2^{13}$CO  (1$_{1,0}$--1$_{1,1}$) was detected in 38 sources  (the entire description is in \S~2 of \citealt{Yan2019}). They found $^{12}$C/$^{13}$C ratio values up to 99 for the source G49.21-0.35, around 32 for G31.41+0.31, and $\sim$~16--47 for W31. 

\underline{$^{16}$O/$^{18}$O and $^{18}$O/$^{17}$O ratios:} The H$_2$C$^{16}$O/H$_2$C$^{18}$O and HC$^{16}$O$^+$/HC$^{18}$O$^+$ ratios provided a rough estimate between $\sim$~170--250. In order to estimate the $^{18}$O/$^{17}$O ratio, from the integrated areas of the lines HC$^{17}$O$^+$ (2--1) and HC$^{18}$O$^+$ (2--1), the ratio HC$^{18}$O$^+$/HC$^{17}$O$^+ \sim$ 3 is obtained, whose value represents an upper limit compared with the abundance ratio HC$^{18}$O$^+$/HC$^{17}$O$^+ \sim$ 1.3, obtained for G331 from the LTE analysis.  In the Galactic center and local ISM, \citet{Wilson1994} obtained ratios of $^{16}$O/$^{18}$O $\sim$~250 and 560, respectively, as well as  $^{18}$O/$^{17}$O $\sim$~3.2 for both of them. \citet{Persson2018} claimed the first detection of H$_2$C$^{17}$O and D$_2^{13}$CO, among other isotopologues, in IRAS 16293-2422 B and reported the values H$_2$C$^{16}$O/H$_2$C$^{18}$O$\sim$ 800, H$_2$C$^{16}$O/H$_2$C$^{17}$O$\sim$ 2596 and H$_2$C$^{18}$O/H$_2$C$^{17}$O$\sim$ 3.2. 

In summary, we present in Table~\ref{tab:tab5} the isotopic ratios estimated for G331  which are in good agreement compared to the values reported from Giant Molecular Clouds (Orion, Sgr B2) and hot molecular cores as G31.41 and G34.26~\citep{Guelin1982,Turner1990,Mangum1990,rob2007,Neill2013,Yan2019}.

\begin{table}
\centering
\caption{Ratios between the isotopologues of H$_2$CO and HCO$^+$ estimated for G331 which are compared with the values reported in   Giant Molecular Clouds (GMCs), i.e., Orion and Sgr (B2), and hot molecular cores (HMCs),  i.e., G31.41 and G34.26.}
\label{tab:tab5}
\centering
\begin{tabular}{l l l l l }
\hline\hline
 Ratios  &  G331 & GMCs  & G31.41 & G34.26 \\
\hline
Formaldehyde & &  \\
H$_2^{13}$CO/H$_2$CO & 0.05 & 0.035$^a$ & 0.03$^e$ & 0.02$^e$ \\
HDCO/H$_2$CO & 0.02 & $\leqslant$ 0.005$^b$ & $<$ 0.002$^f$ & 0.01$^f$ \\
D$_2$CO/HDCO & $\lesssim$0.5 & 0.02$^c$ & --& $<$ 0.001$^f$  \\
H$_2$C$^{18}$O/H$_2$CO & 0.004 & 0.003$^a$ & --& --\\
Formyl cation & &  \\
HCO$^+$/H$^{13}$CO$^+$ & 14 & 21.2$^d$ & -- & -- \\
HC$^{18}$O$^+$/HC$^{17}$O$^+$ & 1.3  & 3.1$^d$ & -- & -- \\
\hline
\end{tabular}\\
a)Orion KL compact ridge \citep{Mangum1990}. b) Orion KL Hot Core \citep{Neill2013}. c) Orion KL Compact Ridge \citep{Turner1990}. d) Sgr B2 \citep{Stark1981,Guelin1982}.  
e) HMCs \citep{Yan2019}. f) HMCs \citep{rob2007}. 
\end{table}


\section{Conclusions and perspectives} \label{sec5}

We have identified isotopologues of formaldehyde, formyl cation and protonated formaldehyde in G331, a hot molecular core and outflow system. The search for those species were carried out using spectral setups    collected with the APEX telescope in the frequency range $\sim$~159--356~GHz. A time dependent chemical model, using the gas grain code {\sc Nautilus}, was performed to study the chemistry and predict the abundances of H$_2$CO, HCO$^+$ and H$_2$COH$^+$ as a function of time. The conclusions and perspectives are:
    
\begin{itemize}

\item Formaldehyde is an abundant molecule in G331. Several lines of H$_2$CO, H$_2^{13}$CO, HDCO and H$_2$C$^{18}$O were detected in G331, 35 spectral lines in total. D$_2$CO was tentatively detected and was used to set an upper threshold for its abundance. From non-LTE calculations, the o- and p-H$_2$CO spectral lines provided kinetic temperatures between $\sim$~87--95~K. From the  population diagrams assuming the LTE approximation, the spectral lines of H$_2^{13}$CO, HDCO and  H$_2$C$^{18}$O provided excitation temperatures of $\sim$~83, 21 and 31~K. From the estimates of column densities considering a source size of 5\arcsec, the ratios H$_2$CO:H$_2^{13}$CO:HDCO:H$_2$C$^{18}$O$\thickapprox$223:12:4:1 were obtained. The formaldehyde isotopologues can trace different physical conditions, from  cold gas to lukewarm temperatures.  These results agree with sublimation processes and the active gas-grain chemistry expected in hot molecular cores.
 
\item The formyl cation is also an abundant species  in G331. In total, 8 rotational lines of HCO$^+$, H$^{13}$CO$^+$, HC$^{18}$O$^+$ and HC$^{17}$O$^+$ were identified in G331. In contrast with formaldehyde, the $^{17}$O isotopologue, HC$^{17}$O$^+$, was detected but not the deuterated one, DCO$^+$. Assuming the LTE approximation, the population diagram of H$^{13}$CO$^+$ provided a temperature and column density $T_{\rm exc} \thickapprox$ 11~K and $N$(H$^{13}$CO$^+$) $\thickapprox$ 2.1 $\times$ 10$^{14}$~cm$^{-2}$,  whose values were used to estimate the column densities of HC$^{18}$O$^+$ and HC$^{17}$O$^+$. From non-LTE calculations, the spectral lines of HCO$^+$ provided $N$(HCO$^+$) $\thickapprox 3 \times$10$^{15}$ cm$^{-2}$ assuming a temperature of about 30~K.
    
    \item The protonated formaldehyde, H$_2$COH$^+$, should be present under interstellar conditions when H$_2$CO and HCO$^+$ are detected (see Eq.~\ref{Eq:1}). In fact, this cation was also detected through four lines, of which one of them is likely blended  with $^{34}$SO$_2$. From the population diagram considering the LTE formalism, it was obtained $T_{\rm exc} \thickapprox$ 25~K and $N$(H$_2$COH$^+$) $\thickapprox$ 1.4 $\times$ 10$^{14}$~cm$^{-2}$ adopting a source size of 5\arcsec. H$_2$CO and HCO$^+$ can produce H$_2$COH$^+$ under interstellar conditions. In addition, these species could play a key role in the formation of complex organic molecules.
    
    \item A gas grain chemical model was performed to predict the fractional abundances of H$_2$CO, HCO$^+$ and H$_2$COH$^+$ and to study their evolution. The best model was obtained adopting $T=$30~K and $n_{\text{H}_2} = 1 \times$ 10$^6$ cm$^{-3}$. In a young molecular stage of $\sim$~10$^3$~yr, the H$_2$CO, HCO$^+$ and H$_2$COH$^+$ abundances reached comparable values to those derived from the observations, [H$_2$CO] = (0.2--2) $\times$10$^{-8}$, [HCO$^+$] = (0.5--4) $\times$10$^{-9}$ and [H$_2$COH$^+$] = (0.2--2) $\times$10$^{-10}$. The reaction between H$_2$CO and HCO$^+$ is one of the major channels to produce H$_2$COH$^+$. On the other hand, it was noticed that dissociative recombination mechanisms with electrons can rapidly destroy HCO$^+$ and H$_2$COH$^+$ affecting their predicted abundances. The results obtained with the chemical modeling of the three molecular species makes us to develop a hypothesis that the evolution stage of G331 is around $10^3$~yr. Further studies in G331 should be carried out to confirm this hypothesis.

   \item From the multi-line analysis of formaldehyde and the formyl cation, new $^{12}$C/$^{13}$C, H/D, $^{16}$O/$^{18}$O and $^{18}$O/$^{17}$O ratios were inferred in G331 and agree with the results of other works. In particular, 
   deuterium was observed in formaldehyde but not in the formyl cation, HDCO and DCO$^+$, respectively. $^{17}$O was observed in the formyl cation but not in formaldehyde, HC$^{17}$O$^+$ and H$_2$C$^{17}$O, respectively. In perspective, along with new observational analysis, gas-grain chemical models might shed light on the molecular processes that lead different isotopic ratios. 
   
\end{itemize}

\begin{acknowledgements}

We sincerely acknowledge the anonymous reviewer for their constructive and careful revision of the manuscript.
EM acknowledges support under the grant "Mar\'{\i}a Zambrano" from the UHU funded by the Spanish Ministry of Universities and the "European Union NextGenerationEU".  This project has also received funding from the European Union’s Horizon 2020 research and innovation program under the Marie Skłodowska-Curie grant agreement No 872081
and from grant PID2019-104002GB-C21 funded by MCIN/AEI/10.13039/501100011033 and, as appropriate, by ‘ERDF A way of making Europe’, by the ‘European Union’ or by the ‘European Union NextGenerationEU/PRTR’. This work is also supported by the Consejer\'{\i}a de Transformaci\'on Econ\'omica, Industria, Conocimiento y Universidades, Junta de Andaluc\'{\i}a and European Regional Development Fund (ERDF 2014-2020) PY2000764. LB gratefully acknowledges support by the ANID BASAL projects AFB17002 and ACE210002. 

\end{acknowledgements}

%

\vspace{5mm}
\facilities{Atacama Pathfinder Experiment (APEX).}


\software{Based on analysis carried out with the CASSIS software \citep{Vastel2015},  
          RADEX \citep{van2007}, 
          Nautilus \citep{wak15}.
          }




\bibliography{bibliography}{}
\bibliographystyle{aasjournal}



\appendix

\section{Update of the internal partition functions of the detected isotopologues of formaldehyde}\label{appendix}

The internal partition functions of the isotopologues of formaldehyde H$_2^{13}$CO, H$_2$C$^{18}$O, HDCO and D$_2$CO detected in this work have been updated. Since the lists of energies, either experimental or theoretical, of the rovibrational levels for these isotopologues are incomplete for computing the internal partition functions for temperatures up to 500~K, the direct sum expression does not reach the convergence.
Therefore, the internal partition functions of the isotopologues of formaldehyde have to be computed using some other tested approximations~\citep{car2019}, e.g., writing  it  in terms of the product of the rotational contribution ($Q_{\rm rot}(T)$) and the harmonic approximation of the vibrational contribution ($Q_{\rm vib}^{\rm harm}(T)$)~\citep{Herzberg} 

\begin{equation}
\label{rvQapprox}
Q_{\rm rv}(T) \approx Q_{\rm rot}(T) \, Q_{\rm vib}^{\rm harm}(T) ~~.
\end{equation}

\noindent where the rotational contribution $Q_{\rm rot}(T)$ is computed as a direct sum (e.g., \citet{Herzberg}) because there are enough rotational energies for the typical temperatures of the ISM,

\begin{equation}
\label{Qrot}
Q_{\rm rot}(T) = \sum_{\rm i}  g_{\rm ns}^{({\rm i})} \, (2 \, J_{\rm i}
+1) \, e^{-{E^{({\rm rot})}_{\rm i} \over k  T}}   ~~.
\end{equation}

\noindent $E^{({\rm rot})}_{\rm i}$ represents the energy for the i-th
rotational state in the ground vibrational state assuming that uniquely the ground electronic state is populated. The rotational energies of the isotopologues of formaldehyde have been taken from the Cologne Database for Molecular Spectroscopy catalog  (CDMS)~\citep{end2016} and from JPL~\citep{pic1998}. 

The nuclear spin degeneracy $g_{\rm ns}^{({\rm i})}$ is
included in the definition of the rotational contribution of the partition function
  (Eq.~\ref{Qrot}) because, in general, the values of  $g_{\rm ns}^{({\rm i})}$ can be different depending on the symmetry of the rotational states. In the case of formaldehyde, the isotopologues H$_2^{13}$CO, H$_2$C$^{18}$O, and D$_2$CO have symmetry ${\cal C}_{2v}(M)$, whose rovibrational states are characterized by the irreducible representations $A_1$, $A_2$, $B_1$ and $B_2$, whereas the monodeuterated isotopologue HDCO has symmetry ${\cal C}_{s}(M)$ and its states are labeled with the irreducible representations $A^{\prime}$ and $A^{\prime\prime}$~\citep{BJbook}. The labelings of the rotational states of the different isotopologues as well as their nuclear spin statistical weights are given in Table~\ref{tab:nuclear-spin}.

Although the validity of the approximation (\ref{rvQapprox}) has been proven suitable for the typical ISM temperatures~\citep{car2019}, a new check is carried out comparing the values of the internal partition function of the main isotopologue of formaldehyde from 2.725~K to 500~K with those calculated as a direct sum of a comprehensive data set for the rovibrational energy levels provided by~\cite{Al-Refaie2015}. In Tab.~\ref{tab-PartF-main}, the values of the harmonic approximation for the vibrational partition function $Q_{\rm vib}^{\rm harm}(T)$ and the internal partition function $Q_{\rm rv}(T)$ (\ref{rvQapprox}) for the main isotopologue are given. $Q_{\rm vib}^{\rm harm}(T)$ is calculated using the experimental fundamental vibrational energies taken from~\cite{Perrin2003,Perrin2006} and the rotational partition function is computed from the JPL rotational energies predicted up to J=99 and K$_a$=25~\citep{pic1998}. The uncertainties of the rovibrational partition function~\citep{car2019}  are obtained considering the uncertainties of all the rotational energies provided in the JPL database~\citep{pic1998} and the experimental uncertainties of the vibrational fundamental energies~\citep{Perrin2003,Perrin2006}. These values are compared with those from CDMS catalog~\citep{end2016} and from \cite{Al-Refaie2015}. 
According to the relative differences between the approximated internal partition function~(\ref{rvQapprox}) and the one calculated more thoroughly~\citep{Al-Refaie2015}, the two results are comparable in the temperature interval from 2.725~K and 500~K. However, the difference with the values provided in the CDMS catalog is of -8.8\% at 500~K and of -0.72\% at 300~K. Therefore, as the approximation (\ref{rvQapprox}) using a harmonic vibrational partition function is acceptable for the interval of temperature typical for the ISM and improves notably the values of JPL and CDMS catalogs for higher temperatures, this is going to be used for the other isotopologues of formaldehyde.

Tab.~\ref{tab-PartF-isotop} evinces the values of the partition function of H$_2^{13}$CO,  H$_2$C$^{18}$O, D$_2$CO, and HDCO calculated in this work. As in Tab.~\ref{tab-PartF-main}, the values of the calculated partition function are compared with those available in CDMS database~\citep{end2016}. For the case of H$_2^{13}$CO, the values and the uncertainties of $Q_{\rm rv}(T)$ are computed using experimental fundamental vibrational energies~\citep{Ng2017,Wohar1991} for the harmonic partition function and rotational energies up to J=99 and K$_a$=25 taken from JPL database~\citep{pic1998}. The highest relative difference between this work and CDMS values is of -0.76\% at T=300~K. Therefore, at this temperature is not relevant to incorporate the vibrational contribution to the partition function. Nevertheless, the vibrational contribution is important at T=500~K increasing the rotational partition function about 9\%.

For the isotopologue H$_2$C$^{18}$O only the values of the rotational partition function are evinced in Tab.~\ref{tab-PartF-isotop}. These values and their uncertainties have been calculated using the CDMS rotational data (up to J=54, K$_a$=16) complemented with those predicted for the excited rotational levels from JPL database (up to J=20, K$_a$=20) when missing in the CDMS catalog. The vibrational partition function could not be calculated because, as far as we know, there are neither fundamental energies nor excited vibrational energies reported in the literature at all. This hindrance could be overcome providing the vibrational fundamental energies calculated by {\it ab initio} or other empirical approaches. Therefore, the partition function in the present work is practically the same reported in CDMS apart from the uncertainties and the new values of the rotational partition function for temperatures from 300~K to 500~K.

The vibrational and the internal partition functions of the double deuterated isotopologue D$_2$CO are presented in Tab.~\ref{tab-PartF-isotop}. The vibrational contribution and its uncertainty have been calculated using the available experimental vibrational fundamental bands~\citep{Perrin1998,Lohilahti2001,Lohilahti2006}. The values and the uncertainties of the rotational partition function are calculated using the CDMS rotational energies (with data up to J=66 and K$_a$=26) complemented with higher excited rotational energy predictions (up to J=60 and K$_a$=60) given in JPL database. At T=300~K the difference of the updated internal partition function with CDMS values is around 2.6\% and this will be around 20\% at T=500~K.

The values and uncertainties of the internal partition function of the monodeuterated isotopologue HDCO as well as the vibrational contribution are also included in  Tab.~\ref{tab-PartF-isotop}. The values of the rotational partition function are calculated using the CDMS rotational energies (up to J=56 and K$_a$=20) complemented with JPL higher excited rotational energy predictions up to J=90 and K$_a$=50. Since the CDMS data are more accurate they were substituted in the JPL predictions to have a more accurate internal partition function. The vibrational partition function is computed using the available experimental and calculated fundamental energies~\citep{Oka1961,Dangoisse1978,Ellsworth2008,Morgan2018}. The experimental uncertainties of the fundamentals $\nu_1$, $\nu_2$, $\nu_4$, and $\nu_6$ measured by Dispersed Fluorescence spectroscopy~\citep{Ellsworth2008} are considered as 2~cm$^{-1}$ according to the widths of the spectral lines whereas the uncertainties of 
$\nu_3$ and $\nu_5$ bands obtained with DVR calculations are assigned with 1.20~cm$^{-1}$.
By comparing with the CDMS data, the relative difference of the updated internal partition function is 1.5\% larger at T=300~K and at least of 14\% at T=500~K.

In general, the update of the internal partition functions incorporate the vibrational contribution as well as the uncertainties and new values from 300~K to 500~K for the four isotopologues H$_2^{13}$CO,  H$_2$C$^{18}$O, D$_2$CO, and HDCO. As supplementary material, their rotational, vibrational and rovibrational partition functions are reported up to T=500 K using a 1 K interval. This update of the partition functions could be relevant for the estimate of the abundances of the four isotopologues of formaldehyde.

\begin{table}[h]
\caption{Nuclear spin statistical weights of the isotopologues of formaldehyde H$_2$CO, H$_2^{13}$CO, H$_2$C$^{18}$O, D$_2$CO and HDCO associated to the rotational states$^a$.}
\label{tab:nuclear-spin}
\begin{center}
{\footnotesize
\begin{tabular}{cccccc}
\hline\hline
Isotopologues &$\Gamma_{\rm rot}$~$^b$ & $K_a$~$^c$ & $K_c$~$^c$ & $g_{\rm ns}^{({\rm i})}$ & Type~$^g$ \\ \hline
H$_2$CO,H$_2$C$^{18}$O      & $A_1$                  & even  & even & 1                       & para \\
             & $A_2$                  & even  & odd & 1                       & para \\
             & $B_1$                  & odd  & odd & 3                       & ortho \\
             & $B_2$                  & odd  & even & 3                       & ortho \\
\hline
H$_2^{13}$CO & $A_1$                  & even  & even & 2~$^d$                       & para \\
             & $A_2$                  & even  & odd & 2~$^d$                       & para \\
             & $B_1$                  & odd  & odd & 6~$^d$                       & ortho \\
             & $B_2$                  & odd  & even & 6~$^d$                       & ortho \\
\hline
D$_2$CO      & $A_1$                  & even  & even & 6~$^e$                      & ortho \\
             & $A_2$                  & even  & odd & 6~$^e$                        & ortho \\
             & $B_1$                  & odd  & odd & 3~$^e$                        & para \\
             & $B_2$                  & odd  & even & 3~$^e$                        & para \\
\hline
HDCO        & $A^{\prime}$            & ---- & even & 6~$^f$                       & ---- \\
             & $A^{\prime\prime}$     & ---- & odd & 6~$^f$                       & ---- \\
\hline
\hline
\end{tabular}

\begin{flushleft}
$^a$ The  nuclear spin degeneracy $g_{\rm ns}^{({\rm i})}$ is computed according to \cite{BJbook}.  

$^b$ The symmetry labeling of the rotational states.

$^c$ The rotational states of the asymmetric top, such as formaldehyde, are labeled by the quantum numbers $J_{K_a,K_c}$, where $J$ is the rotational angular momentum and the $K_a$ and $K_c$ are the projections of the rotational angular momentum along the $a$- and $c$- molecule-fixed axes. In this case, the symmetry of the rotational states is characterized by their even and odd values of $K_a$ and $K_c$. For the monodeuterated isotopologue, the two symmetries are only characterized by the even and odd values of $K_c$. 

$^d$ For H$_2^{13}$CO, the nuclear spin degeneracy is also considered in the literature with a ratio 3:1 for the ortho:para states, respectively (see, e.g., \cite{end2016}). As a warning in order to avoid wrong results, before using data from a catalog, it should be checked whether the nuclear spin weights agree with the partition function considered.

$^e$ For D$_2$CO, the nuclear spin degeneracy is also considered in the literature with a ratio 2:1 (ortho:para) (see, e.g., \cite{end2016}). Same warning from footnote $d$ should be considered in this case.

$^f$ For the monodeuterated isotopologue, the nuclear spin degeneracy is considered in this paper as 1 because the degeneracy is state independent.

$^g$ Only for the symmetric isotopologues of formaldehyde, this column shows whether the transitions involving these rotational states are either ortho or para. 

\end{flushleft}
}
\end{center}
\end{table}

\begin{table}
\caption{Vibrational and rotational-vibrational partition function of the main isotopologue of formaldehyde (H$_2$CO). Comparison between the values obtained in the present study and those published before.$^a$}
\label{tab-PartF-main}
\begin{center}
{\footnotesize
\begin{tabular}{c|cr|rr|c}
\hline\hline
$T$(K) &$Q_{\rm vib}^{\rm harm}$~$^b$ & $Q_{\rm rv}$(Present work)$^c$ & $Q$(CDMS)$^d$ & $Q$\citep{Al-Refaie2015}$^e$ & Rel. Diff.(\%)$^f$ \\ \hline
    2.725& 1.000000 &      2.0166(0) &    2.0166 &    2.0165 & 0.00\\
    5.000& 1.000000 &      4.4832(0) &    4.4832 &    4.4833 & 0.00\\
    9.375& 1.000000 &     13.8009(0) &   13.8008 &   13.8010 & 0.00\\
   18.750& 1.000000 &     44.6813(0) &   44.6812 &   44.6835 & 0.00\\
   37.500& 1.000000 &    128.6496(0) &  128.6492 &  128.6581 & 0.01\\
   75.000& 1.000000 &    361.7207(0) &  361.7195 &  361.7053 & 0.00\\
  150.000& 1.000021 &   1019.9947(0) & 1019.9706 & 1019.9549 & 0.00\\
  225.000& 1.000996 &   1874.4927(0) & 1872.6221 & 1874.4679 & 0.00\\
  300.000& 1.007228 &   2903.8609(0) & 2883.0163 & 2904.1778 & 0.01\\
  500.000& 1.087521 &   6751.7086(3) & 6208.3442 & 6760.2315 & 0.13\\
\hline
\hline
\end{tabular}

\begin{flushleft}
$^a$ The  nuclear spin degeneracy is given in Tab.\ref{tab:nuclear-spin}.

$^b$ Values of the vibrational partition function computed with the harmonic approximation. For more details, see the text.

$^c$ $Q_{\rm rv}=Q_{\rm rot}(\mbox{\rm direct sum}) \, Q_{\rm
    vib}^{\rm harm}$. An estimate of the uncertainties is given in parentheses in  units of the last quoted digits. For more details, see the text. 

$^d$ Rotational partition function computed as a direct sum with no vibrational contribution. Their values are reported in CDMS catalog~\citep{end2016}.

$^e$ Internal partition function computed as the direct sum using a comprehensive set of rovibrational energies up to 18000~cm$^{-1}$ and up to J=70~\citep{Al-Refaie2015}. 

$^f$ Relative difference of the partition function computed in the present study with respect to the one reported by \cite{Al-Refaie2015}.

\end{flushleft}
}
\end{center}
\end{table}

\begin{table}
\caption{Vibrational and rotational-vibrational partition functions for the isotopologues H$_2^{13}$CO, H$_2$C$^{18}$O, D$_2$CO and HDCO. Comparison between the values
  obtained in the present study and those published in CDMS catalog$^a$.}
\label{tab-PartF-isotop}
\begin{center}
{\footnotesize
\begin{tabular}{c|cccc|cccc}
\hline\hline
&   \multicolumn{4}{c}{H$_2^{13}$CO} &   \multicolumn{4}{c}{H$_2$C$^{18}$O}  \\ 
\hline
$T$(K) &$Q_{\rm vib}^{\rm harm}$~$^b$ & $Q_{\rm rv}$(Present work)$^c$ & $Q$(CDMS)$^d$ & Rel. Diff.(\%)$^e$ &\multicolumn{2}{c}{$Q_{\rm rot}$(Present work)$^{c}$} & $Q$(CDMS)$^d$ & Rel. Diff.(\%)$^e$ \\ \hline
  2.725&1.000000  &    4.1136(0) &    4.1136 & 0.00 & \multicolumn{2}{c}{2.0944(0)}    & 2.0944 & 0.00\\
  5.0  &1.000000  &    9.1701(0) &    9.1700 & 0.00 & \multicolumn{2}{c}{4.6806(0)}    & 4.6805 & 0.00\\
  9.375&1.000000  &   28.2638(0) &   28.2636 & 0.00 & \multicolumn{2}{c}{14.4425(0)}   & 14.4424 &0.00\\
 18.75 &1.000000  &   91.5675(0) &   91.5672 & 0.00 & \multicolumn{2}{c}{46.8159(0)}   & 46.8156 &0.00\\
 37.50 &1.000000  &  263.7391(0) &  263.7382 & 0.00 & \multicolumn{2}{c}{134.8810(0)}  & 134.8805 &0.00 \\
 75.0  &1.000000  &  741.6749(0) &  741.6726 & 0.00 & \multicolumn{2}{c}{379.3628(0)}  &  379.3616 &0.00  \\
150.0  &1.000023  & 2091.5874(3) & 2091.5330 & 0.00 & \multicolumn{2}{c}{1069.8884(0)} & 1069.8850  &0.00  \\ 
225.0  &1.001065  & 3844.176(18) & 3840.0722 &-0.11 & \multicolumn{2}{c}{1964.3668(0)} & 1964.3606  & 0.00 \\
300.0  &1.007605  & 5957.09(14)  & 5912.1084 &-0.76 & \multicolumn{2}{c}{3024.3241(1)} & 3024.3231  &  0.00\\
500.0  &1.090530  &13884.1(19)   &   ----    & ---- & \multicolumn{2}{c}{6508.862(13)} & ----  & ---- \\
\hline
\hline
&   \multicolumn{4}{c}{D$_2$CO} &   \multicolumn{4}{c}{HDCO}  \\ 
\hline
$T$(K) &$Q_{\rm vib}^{\rm harm}$~$^b$ & $Q_{\rm rv}$(Present work)$^c$ & $Q$(CDMS)$^d$ & Rel. Diff.(\%)$^e$ &$Q_{\rm vib}^{\rm harm}$~$^b$ & $Q_{\rm rv}$(Present work)$^{c}$ & $Q$(CDMS)$^d$ & Rel. Diff.(\%)$^e$ \\ \hline
    2.725 & 1.000000     & 14.984964(0)     &14.9850  &0.00 &1.000000    &  2.2608(0)     &2.2608  & 0.00 \\
    5.000 & 1.000000     & 29.703636(0)     &29.7036  &0.00 &1.000000    &  4.6483(0)     &4.6483  & 0.00 \\
    9.375 & 1.000000     & 67.339744(0)    & 67.3395 & 0.00 &1.000000    &  11.2681(0)    &11.2680  &0.00  \\
   18.750 & 1.000000     & 182.210324(0)    &182.2098  &0.00  &1.000000    & 31.0556(0)     &31.0555  & 0.00 \\
   37.500 & 1.000000     &  509.265756(0)   & 509.2641 &0.00  &1.000000    & 86.7359(0)    &86.7356  &0.00  \\
   75.000 & 1.000000     & 1432.775532(0)   & 1432.7709 &0.00  &1.000000    & 243.8496(0)    & 243.8488 &0.00  \\
  150.000 & 1.000226     & 4044.511598(0)   & 4043.5866 &0.02  &1.000093    &  688.0292(0)  & 687.9631 &-0.01  \\
  225.000 & 1.005192     & 7464.6706(1)   & 7426.0941 &-0.52  &1.002701    & 1266.752(29)   & 1263.3349 &-0.27  \\
  300.000 & 1.025736     & 11729.1102(6)  & 11434.7820 &-2.57 &1.015159    & 1974.73(18)  &1945.2333  &-1.52 \\ 
  500.000 & 1.202824     & 29624.2096(74)  & ----  & ----     &1.138623    & 4769.8(21) &  ---- & ---- \\
\hline\hline
\end{tabular}

\begin{flushleft}
$^a$ The  nuclear spin degeneracy is considered according to Tab.~\ref{tab:nuclear-spin}.

$^b$ The vibrational partition function is  computed with the harmonic approximation. For more details, see the text.

$^c$ $Q_{\rm rv}=Q_{\rm rot}(\mbox{\rm direct sum}) \, Q_{\rm
    vib}^{\rm harm}$. An upward estimate of the uncertainties is given in parentheses in  units of the last quoted digits. For more details, see the text. For the isotopologue H$_2$C$^{18}$O, it is presented the rotational partition function because there are neither experimental nor theoretical  vibrational levels available in the literature.

$^d$ Rotational partition function computed as a direct sum with no vibrational contribution. Their values are reported in CDMS catalog~\citep{end2016}. The CDMS partition functions of H$_2^{13}$CO and D$_2$CO are multiplied by 2 and 3, respectively, to consider the same nuclear spin statistical weight from Tab.~\ref{tab:nuclear-spin}.

$^e$ Relative difference of the partition function given in the present study with respect to the one reported in CDMS catalog.

\end{flushleft}
}
\end{center}
\end{table}

\end{document}